\documentclass[preprint,10pt]{elsarticle}

\usepackage{amsmath,amssymb,amsfonts} 
\usepackage{bm}
\usepackage{graphicx} 
\usepackage{textcomp} 
\usepackage{xcolor} 
\usepackage{listings} 
\usepackage{algorithm} 
\usepackage{multirow} 
\usepackage{algpseudocode}
\usepackage{subcaption}
\usepackage{booktabs} 
\usepackage[hyphens]{url} 
\usepackage{fancyhdr} 
\usepackage{xcolor} 
\usepackage{indentfirst}
\usepackage{resizegather}
\usepackage{colortbl} 
\usepackage{makecell}
\usepackage{array} 
\usepackage{enumitem} 
\usepackage{hhline}
\usepackage{tikz}
\usepackage{caption} 
\usepackage[margin=1in]{geometry}
\usepackage{comment}
\usepackage{color}
\usepackage{hyperref}
\hypersetup{colorlinks=true,linktoc=all,linkcolor=blue,pdfauthor={Name}}
\usepackage{gensymb}
\usepackage{rotating}
\usepackage{wrapfig}

\usepackage{mathrsfs}
\usepackage{float}
\usepackage[nameinlink]{cleveref}

\usepackage{booktabs}
\usepackage{tabularx}
\usepackage{array}
\usepackage{threeparttable}

\usepackage{longtable}
\usepackage{microtype}
\usepackage{xcolor}
\usepackage{listings}
\usepackage{siunitx}
\usepackage{lineno}

\usepackage{tikz}
\usetikzlibrary{arrows.meta,positioning,calc,fit,shapes.geometric}

\hypersetup{
  colorlinks=true,
  linkcolor=blue!55!black,
  citecolor=blue!55!black,
  urlcolor=blue!55!black
}

\definecolor{codebg}{RGB}{246,247,249}
\definecolor{codeframe}{RGB}{210,214,220}
\definecolor{d01c}{RGB}{225,238,255}
\definecolor{d02c}{RGB}{215,245,225}
\definecolor{d03c}{RGB}{255,239,204}
\definecolor{d04c}{RGB}{255,218,218}
\definecolor{obsc}{RGB}{170,20,20}
\definecolor{feedc}{RGB}{25,90,160}

\lstdefinelanguage{WRFNamelist}{
  morecomment=[l]{!},
  morestring=[b]',
  morestring=[b]",
  sensitive=false
}
\journal{Atmosphere}

\newcommand{\code}[1]{\texttt{#1}}

\newcommand{\sgs}{\mathrm{SGS}}

\begin{document}
\begin{frontmatter}

\title{Site-scale, Multi-resolution WRF Wind Modeling with Observational Nudging for Methane Emission Monitoring in the Permian Basin}

\author[MIT,Brown]{Ehsan~Kharazmi\corref{cor1}}
\ead{kharazmi@mit.edu}

\author[MIT]{Erin~Menezes}
\ead{emenezes@mit.edu}


\author[MIT,Brown]{George~Karniadakis}
\ead{george_karniadakis@brown.edu}

\author[EM]{Anantha~Sundaram}
\ead{anantha.sundaram@exxonmobil.com}

\author[EM]{Arash~Fathi}
\ead{arash.fathi@exxonmobil.com}


\cortext[cor1]{Corresponding author}

\affiliation[MIT]{organization={Massachusetts Institute of Technology}, city={Boston}, state={MA}, country={USA}}

\affiliation[Brown]{organization={Brown University}, city={Providence}, state={RI}, country={USA}}

\affiliation[EM]{organization={ExxonMobil Technology and Engineering Company}, city={Spring}, state={TX}, country={USA}}

\begin{abstract}
Accurate methane source localization and emission-rate estimation at oil and gas facilities require wind fields that resolve site-scale spatial and temporal variability, which cannot be characterized by a single anemometer or coarse operational weather products. We develop a multiscale Weather Research and Forecasting (WRF) framework for a Permian Basin facility that dynamically downscales hourly, $3~\mathrm{km}$ High-Resolution Rapid Refresh (HRRR) fields through four nested domains with grid spacings of $3~\mathrm{km}$, $1~\mathrm{km}$, $200~\mathrm{m}$, and $40~\mathrm{m}$. The two outer domains use planetary-boundary-layer parameterization, whereas the two inner domains operate in large-eddy-simulation mode. One-minute wind observations from two on-site anemometers are assimilated through wind-only observational nudging in the innermost domain, and the effects of nesting feedback and turbulence-closure choices are examined. Simulations for winter and summer 2025 periods are evaluated using near-surface wind-speed time series, power spectra, spatial fields, and statistical metrics. The baseline simulation reproduces the broad evolution of observed wind events but drifts during weak-wind periods. Observational nudging with one- and two-way nesting reduces the absolute bias of wind-speed by \%65 at one sensor and \%87 at the other,
while reducing the corresponding root-mean-square errors by \%33 and \%44.
Nudging also increases high-frequency energy and resolved spatial gradients, although improvements in turbulence statistics are not uniform. These results demonstrate a practical physics-based pathway for generating high-resolution wind fields for methane-plume modeling while emphasizing the need for independent spatial observations to validate accuracy away from assimilated sensors.
\end{abstract}


\end{frontmatter}

\begingroup
\hypersetup{linkcolor=black}
\endgroup

\section{Introduction}
\noindent
Continuous methane monitoring is increasingly used at oil and gas (O\&G) facilities to detect intermittent releases and estimate their source locations and emission rates \cite{atmos13040510,Fathi2023AGU}. The reliability of these estimates depends not only on the methane concentration measurements, but also on the wind field used to transport the simulated plume between a candidate source and the sensor network. In practice, a facility may contain several methane sensors but only one or a few anemometers. Such point measurements cannot fully describe the three-dimensional, time-evolving wind velocity field across a site, particularly when buildings, equipment, surface heterogeneity, boundary-layer shear, and turbulence produce variations over tens to hundreds of meters. Conversely, operational meteorological products such as the High-Resolution Rapid Refresh (HRRR) provide physically consistent regional fields but, at approximately $3~\mathrm{km}$ horizontal and hourly resolution, do not directly resolve the flow structures relevant to a facility with dimensions of only a few hundred meters. Bridging this mismatch is therefore a central meteorological challenge for physics-based methane-plume modeling and inverse emission estimation.

Dynamical downscaling with the Weather Research and Forecasting (WRF) model offers a framework for connecting synoptic and mesoscale forcing to microscale atmospheric flow. Its Advanced Research WRF dynamical core integrates the fully compressible, nonhydrostatic equations using a time-split formulation \cite{SkamarockKlemp2008}. In a nested multiscale configuration, outer domains retain the evolving regional weather and parameterize planetary-boundary-layer turbulence, while progressively finer domains explicitly resolve an increasing fraction of the energetic eddies through large-eddy simulation (LES). Early real-case studies demonstrated both the promise and the limitations of this approach. Talbot et al. \cite{Talbot2012} nested WRF from mesoscale resolution to $50~\mathrm{m}$ and found that fine grids improved the representation of surface heterogeneity and some near-surface conditions, but that errors in the driving meteorology propagated into the LES domains and prevented uniform improvement across all variables. Bhimireddy and Bhaganagar \cite{Bhimireddy2018} dynamically downscaled from $24~\mathrm{km}$ to $150~\mathrm{m}$ for atmospheric-dispersion applications and showed that the choice of forcing data, boundary-layer physics, vertical resolution, and nudging affected the resolved flow and turbulence. These findings illustrate that increasing resolution alone is insufficient; a credible site-scale solution also requires a controlled transition in physics across the mesoscale, turbulence gray zone, and LES regimes.

Recent studies over complex terrain have provided a more detailed assessment of these modeling choices. Liu et al. \cite{Liu2020} showed that WRF--LES at $37~\mathrm{m}$ could reproduce both large-scale events and microscale circulation features, with high-resolution topography and the treatment of turbulence strongly affecting near-surface wind statistics. Using an ensemble of 36 multiscale WRF configurations extending from $11.25~\mathrm{km}$ to $30~\mathrm{m}$, Giani and Crippa \cite{GianiCrippa2024} found that simulated winds were most sensitive to terrain and land-use data, with important contributions from the large-scale forcing and gray-zone turbulence treatment but comparatively weak sensitivity to the LES subgrid-scale closure. Al Oqaily et al. \cite{AlOqaily2025} subsequently evaluated that ensemble against flux-tower and radiosonde observations from the Perdig\~ao field campaign. Their results showed that LES and detailed surface data improved the representation of recirculation and other terrain-induced structures, but did not consistently reduce pointwise wind-speed errors; near-surface performance depended on location, time of day, synoptic regime, boundary forcing, and gray-zone physics. Janiszeski and Crippa \cite{JaniszeskiCrippa2025} similarly used multiscale WRF simulations extending to $30~\mathrm{m}$ to reveal the microscale structure and turbulence of Sundowner wind events that were not captured at mesoscale resolution, while noting that improvements in pointwise statistical metrics were comparatively modest. Together, these studies establish an important distinction between resolving physically meaningful spatial structure and obtaining better agreement at an individual observation point.

The mesoscale-to-microscale transition introduces additional numerical and physical challenges. The gray zone is too fine for conventional one-dimensional boundary-layer parameterizations to remain fully valid, yet too coarse for the dominant turbulent motions to be adequately resolved by LES. Wyngaard \cite{Wyngaard2004} characterized this regime as the ``terra incognita,'' in which the filter scale becomes comparable to the scale of the energy-containing turbulent motions. Smooth inflow inherited from a mesoscale parent can also require substantial fetch and spin-up before realistic turbulence develops in a child domain. Mu\~noz-Esparza et al.\ \cite{MunozEsparza2014} showed that switching to three-dimensional LES mixing does not guarantee immediate turbulence development and demonstrated that inflow temperature perturbations can accelerate the transition. Even in LES-to-LES nesting, Mirocha et al.\ \cite{Mirocha2013} found near-inflow deficits in wind speed and momentum flux, with sensitivity to subfilter-scale closure, mesh resolution, and grid aspect ratio.

In a controlled convective case over the flat SWiFT site, the FY2019 DOE Mesoscale-to-Microscale Coupling report found that turbulence without inflow perturbations required approximately $5~\mathrm{km}$ of fetch merely to initiate, whereas stochastic cell perturbations and synthetic-inflow methods substantially accelerated its development \cite{Haupt2019MMC}. Because the required fetch depends on stability, surface heterogeneity, forcing, grid resolution, and the perturbation method, that result motivates explicit assessment of turbulence development rather than prescribing a universal fetch requirement. The subsequent review by Haupt et al. \cite{Haupt2023} identifies gray-zone treatment, turbulence initialization, surface heterogeneity, coupling strategy, and application-specific validation metrics as central unresolved issues in mesoscale-to-microscale modeling. Grid design is also important: Daniels et al. \cite{Daniels2016} demonstrated that horizontal and vertical refinement should be selected with attention to grid aspect ratio and terrain-following-coordinate errors. These results motivate a nesting hierarchy that changes the turbulence treatment deliberately across scales and evaluates not only mean wind error, but also temporal spectra, spatial variability, shear, and coherent flow structure.

Data assimilation provides a complementary way to constrain dynamically downscaled fields with observations. At regional scales, grid and spectral nudging have been used to prevent a limited-area simulation from drifting away from its large-scale forcing while retaining smaller-scale variability \cite{LiuNenes2012}. Sensitivity experiments also indicate that horizontal wind is the most influential variable to nudge when the objective is to maintain a dynamically consistent regional circulation \cite{Omrani2015}. Observation nudging instead relaxes the modeled state toward individual measurements within prescribed spatial and temporal influence windows, making it attractive when observations are sparse. Stauffer and Seaman \cite{StaufferSeaman1994} established the importance of matching the scales of the assimilated information to the nested model hierarchy: coarse-grid analysis nudging and finer-grid observation nudging served complementary roles in their mesoscale experiments. Applications have shown improved regional wind fields when terrain-aware spatial correlations are included \cite{Ren2019} and when Doppler-lidar wind profiles are assimilated over an urban region \cite{NayakKanda2023}. Sommerfeld et al.\ \cite{Sommerfeld2019} also found improved winds at a lidar measurement location, but the response varied with height and was limited close to the surface. Using a dense surface network and $4~\mathrm{km}$ WRF simulations, Yi et al.\ \cite{Yi2020} demonstrated improvements in surface variables against independent observations, while precipitation did not improve further when surface observation nudging was added to analysis nudging. For the DISCOVER-AQ Texas campaign, Li et al.\ \cite{Li2016Nudging} showed that improved meteorological fields could benefit simulated ozone concentrations, although some local wind shifts remained unresolved. These studies motivate observational constraints for transport applications without implying uniformly improved skill across variables or scales.

More recently, Jiang et al. \cite{Jiang2026} combined observation nudging with an $88.9~\mathrm{m}$ WRF--LES domain and reported improved wind and turbulence statistics at an offshore measurement location. Nevertheless, applying near-surface observations directly within a turbulence-resolving domain remains delicate: an overly strong constraint can alter the turbulence that LES is intended to generate, and agreement at an assimilated sensor is not an independent validation of the surrounding wind field. Moreover, most prior applications have focused on wind energy, complex terrain, or urban meteorology. Comparatively little work has examined how sparse, near-surface anemometer measurements constrain a WRF--LES domain with grid spacing of only a few tens of meters at an industrial facility, or how the resulting correction interacts with nested-domain feedback.

In this study, we develop and evaluate a multiscale WRF framework for 
a representative O\&G site in the Permian Basin. Hourly HRRR fields at $3~\mathrm{km}$ resolution provide the initial and lateral boundary conditions for four nested WRF domains with horizontal grid spacings of $3~\mathrm{km}$, $1~\mathrm{km}$, $200~\mathrm{m}$, and $40~\mathrm{m}$. The two outer domains use planetary-boundary-layer parameterization, whereas the two inner domains operate in LES mode, providing a controlled transition from regional forcing to site-scale resolved flow. One-minute measurements from two anemometers mounted $1.5~\mathrm{m}$ above ground are assimilated through wind-only observational nudging applied in the innermost domain. We additionally examine the roles of one- and two-way nesting and alternative subfilter-scale configurations. Winter and summer  periods from 2025 are assessed using near-surface time series, error and correlation metrics, power spectral density, spatial wind fields, and domain-scale measures of variability. Because both anemometers are used in the nudging procedure, comparisons at those locations are interpreted as measures of the assimilation response rather than as independent spatial validation. Preliminary results of this framework were presented in \cite{Fathi2025AGUWind}.

The principal contribution of this work is therefore not spatial resolution alone, but the integration and critical evaluation of regional HRRR forcing, a mesoscale-to-LES nesting hierarchy, and sparse near-surface observational nudging for methane-monitoring applications. The analysis addresses three related questions: whether the nested configuration adds site-scale spatial and temporal structure beyond the HRRR forcing, how observational nudging changes near-surface errors and resolved variability, and how sensitive those changes are to nesting feedback and subfilter-scale treatment. The resulting fields provide physically evolving wind information at scales relevant to facility-scale plume transport, while the evaluation identifies which improvements are robust and which aspects still require independent observations. The simulations also provide a physics-based reference for the future development of computationally efficient surrogate or scientific machine-learning models for operational wind-field generation \cite{Fathi2025AGU}.

The remainder of the paper is organized as follows. \Cref{sec:problem_description} to \Cref{sec:field sensor} describe the site, observations, nested-domain configuration, and turbulence treatment. \Cref{sec:nudging} presents the observational-nudging implementation and the role of two-way nesting. \Cref{sec:experiments} evaluates the baseline and nudged simulations across the selected meteorological periods and sensitivity cases. Finally, \Cref{sec:Conclusions} summarizes the findings, limitations, and implications for methane-plume modeling and emission monitoring.

\section{Problem Description, Observations, and Model Configuration} \label{sec:problem_description}

Building on the motivation established in the Introduction, we focus on 
a representative O\&G site
in the Permian Basin. The monitored portion of the site occupies approximately $300~\mathrm{m}\times300~\mathrm{m}$ and contains methane-concentration sensors, two near-surface anemometers, storage tanks, and other production equipment. The available observations therefore provide detailed temporal information at only a few points, rather than a complete three-dimensional wind field across the facility.

We use the High-Resolution Rapid Refresh (HRRR) product to supply the evolving regional atmospheric state. HRRR fields at approximately $3~\mathrm{km}$ horizontal and hourly temporal resolution are processed through the WRF Preprocessing System (WPS) and \texttt{real.exe} to generate the initial and lateral boundary conditions for WRF. In this role, HRRR provides the synoptic and mesoscale forcing; it is not treated as a direct representation of the facility-scale wind.

WRF dynamically downscales this forcing through four nested domains of progressively finer resolution, with the innermost domain centered on the facility and reaching $40~\mathrm{m}$ horizontal grid spacing. The model evolves wind, temperature, moisture, and boundary-layer structure consistently across the domain hierarchy. The resulting fields are used for facility-scale plume-transport applications and as physics-based reference data for future scientific machine-learning models, while the available anemometer measurements provide the local observational constraint described in \Cref{sec:nudging}.

\section{Our WRF Setup}

\begin{figure}[h]
    \centering
    \includegraphics[width=0.8\linewidth]{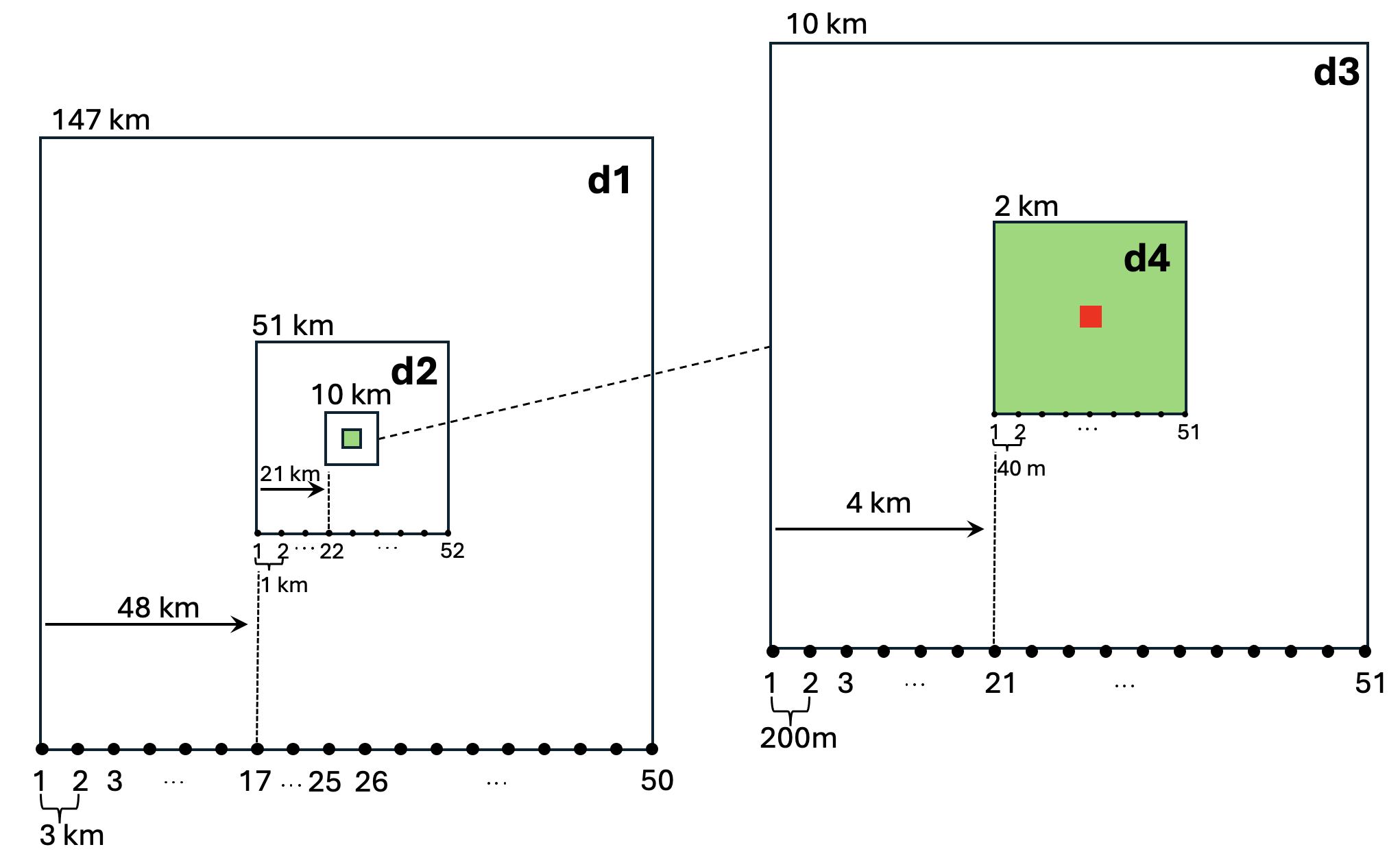}
    \includegraphics[width=0.7\linewidth]{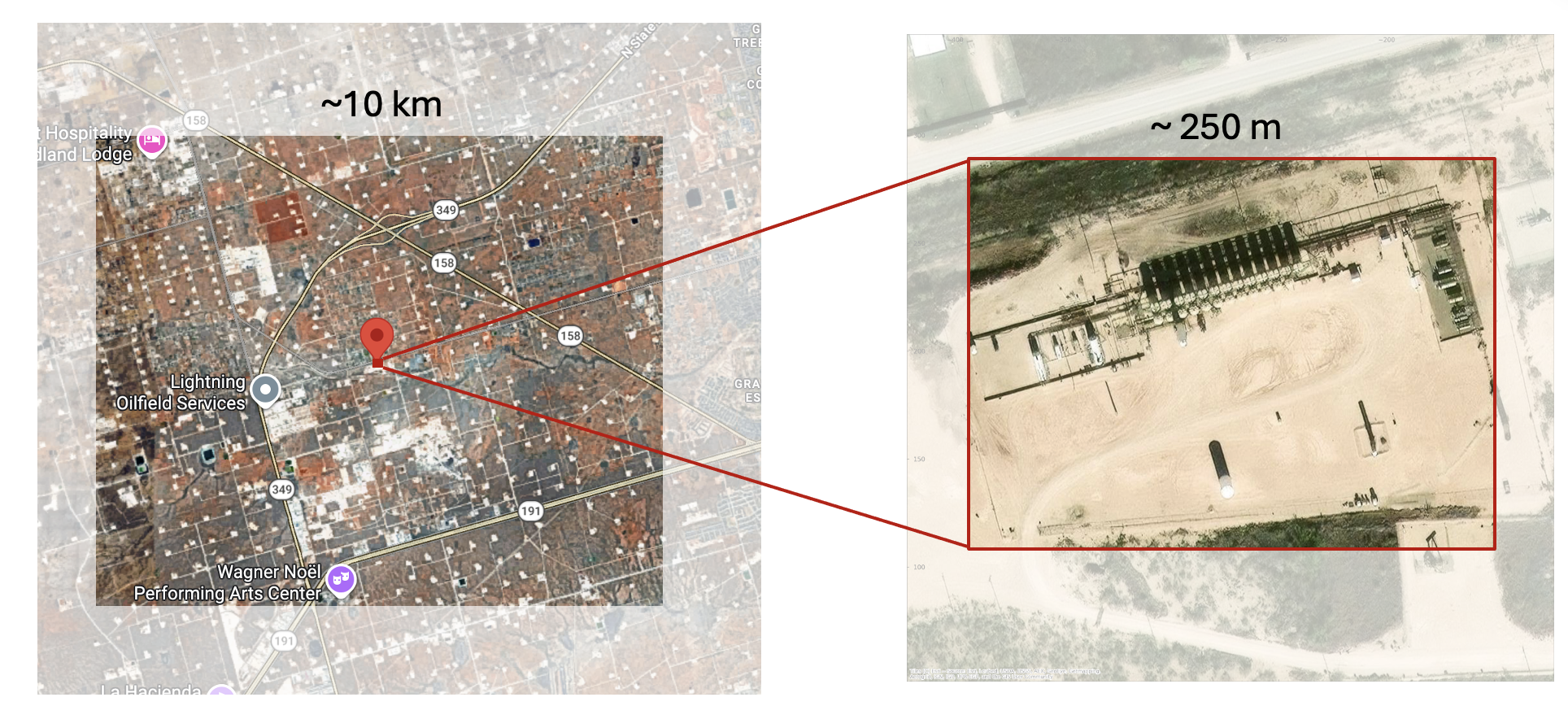}
    \caption{Top: Schematic of the nested domains in WRF simulation setup. Bottom: Satellite image of the O\&G site location in two scales of 9 km and 200 m. The site fits in a domain with 200-300 m side length. }
    \label{fig:domains}
\end{figure}

Bridging the gap between the $3~\mathrm{km}$ HRRR forcing and the flow scales relevant to facility-scale plume transport requires a model configuration that preserves the large-scale atmospheric evolution while progressively resolving smaller-scale motions near the site. We therefore use four nested WRF domains, shown in \Cref{fig:domains} and summarized in \Cref{tab:wps_domains}. The outermost domain (d01) matches the horizontal resolution of the HRRR initial and lateral boundary conditions, while successive parent-to-child refinement ratios of 3, 5, and 5, selected following WRF nesting best practices \cite{wrf_wps_best_practices}, reduce the grid spacing to $40~\mathrm{m}$ in d04. This hierarchy is designed not only to increase spatial resolution, but also to provide a controlled transition from mesoscale, PBL-parameterized modeling in d01--d02 to turbulence-resolving LES in d03--d04. The remainder of this section describes the numerical hierarchy, the change in turbulence treatment across scales, the spin-up strategy, and the computational requirements of the simulations.


\begin{table*}[t]
\centering
\caption{WRF computational-domain configuration.}
\label{tab:wps_domains}
\begin{threeparttable}
\small
\setlength{\tabcolsep}{6pt}
\renewcommand{\arraystretch}{1.15}
\begin{tabularx}{\textwidth}{
@{}
>{\raggedright\arraybackslash}X
>{\centering\arraybackslash}p{0.15\textwidth}
>{\centering\arraybackslash}p{0.15\textwidth}
>{\centering\arraybackslash}p{0.15\textwidth}
>{\centering\arraybackslash}p{0.15\textwidth}
@{}
}
\toprule
\textbf{Horizontal grid and nesting}
& \textbf{d01}
& \textbf{d02}
& \textbf{d03}
& \textbf{d04} \\
\midrule
Domain role
& Outer domain
& Nested domain
& Nested domain
& Innermost domain \\
Parent domain
& --
& d01
& d02
& d03 \\
Parent-grid ratio
& 1
& 3
& 5
& 5 \\
Horizontal grid spacing, $\Delta x=\Delta y$
& 3 km
& 1 km
& 200 m
& 40 m \\
Grid dimensions,
$e_{\mathrm{we}}\times e_{\mathrm{sn}}$
& $50\times50$
& $52\times52$
& $51\times51$
& $51\times51$ \\
Approximate horizontal extent
& $147\times147$ km$^2$
& $51\times51$ km$^2$
& $10\times10$ km$^2$
& $2\times2$ km$^2$ \\
Parent-domain start index,
$(i_{\mathrm{start}},j_{\mathrm{start}})$
& $(1,1)$
& $(17,17)$
& $(22,22)$
& $(21,21)$ \\ \midrule
\textbf{Physics and Dynamics}
& \textbf{d01}
& \textbf{d02}
& \textbf{d03}
& \textbf{d04} \\ \midrule
Effective model time step
& 5.000 s
& 1.667 s
& 0.333 s
& 0.067 s \\
Vertical coordinate
& \multicolumn{4}{c}{
Hybrid sigma--pressure coordinate (2)
} \\
Vertical grid and model top
& \multicolumn{4}{c}{
54 eta levels; model top at 5000 Pa
} \\
Turbulent mixing (Vertical)
& YSU PBL
& YSU PBL
& 1.5-order TKE
& 1.5-order TKE \\
Eddy-coefficient/turbulence closure (Horizontal)
& 2-D Smagorinsky
& 2-D Smagorinsky
& 1.5-order TKE
& 1.5-order TKE \\
\bottomrule
\end{tabularx}
\end{threeparttable}
\end{table*}

Together, the grid hierarchy and domain-specific time steps provide the numerical framework for carrying synoptic forcing from HRRR into the site-scale domain. Domains d01 and d02 represent the mesoscale evolution of the atmosphere, whereas d03 and d04 progressively admit explicitly resolved turbulent motions. We employ both one-way and two-way nesting options. In one-way nesting, each parent supplies time-dependent lateral boundary conditions to its child, while the child does not modify the parent. In two-way nesting, the child domain will feed back to the parent domain. The description of our two-way nesting with observation nudging is explained in detail in Section \ref{sec:nudging}. 


\subsection{Transition from mesoscale modeling to LES}

The nested configuration transitions from mesoscale dynamics to turbulence-resolving flow in a manner consistent with the changing separation between resolved and unresolved atmospheric motions. In the two outer domains (d01--d02; $3000$--$1000~\mathrm{m}$ grid spacing), boundary-layer turbulence is primarily parameterized using the Yonsei University (YSU) PBL scheme (\texttt{bl\_pbl\_physics=1}), which represents the net effect of unresolved eddies on the resolved mean flow through a nonlocal K-profile closure with explicit entrainment treatment \cite{Hong2006}. At these resolutions, the dominant energy-containing eddies responsible for momentum and scalar transport remain largely subgrid, so the turbulent fluxes governing near-surface winds and mixing must be supplied by the PBL parameterization and associated diffusion operators. Horizontal mixing is represented with a Smagorinsky-type formulation in d01 and d02 (\texttt{km\_opt=4}), with diffusion evaluated in physical space (\texttt{diff\_opt=2}) to relate the mixing more directly to local flow gradients.

In the inner domains (d03--d04; $200$--$40~\mathrm{m}$ grid spacing), the PBL scheme is disabled (\texttt{bl\_pbl\_physics=0}) and the model is operated in LES mode. A substantial fraction of the turbulent spectrum---particularly the larger eddies controlling short-time wind variability and dispersion---is then resolved explicitly by the prognostic momentum equations. Subfilter turbulence is represented using a 1.5-order TKE-based subgrid-scale (SGS) closure (\texttt{km\_opt=2}) together with a TKE-based subfilter-stress option (\texttt{sfs\_opt=2}), with diffusion evaluated in physical space (\texttt{diff\_opt=2}). Turbulent kinetic energy is consequently partitioned between resolved wind fluctuations and a modeled SGS component that supplies the stresses and dissipation missing near the filter scale. This shift from predominantly parameterized turbulence in the mesoscale domains to increasingly resolved turbulence in the LES domains is central to representing the intermittency, directional meander, and shear-driven fluctuations that affect plume transport at the site.

Surface forcing is kept consistent across the domain hierarchy through the revised MM5 similarity surface-layer scheme (\texttt{sf\_sfclay\_physics=1}) coupled to Noah-MP land-surface physics (\texttt{sf\_surface\_physics=4}). These schemes provide the surface fluxes that drive boundary-layer evolution in d01--d02 and turbulence generation in d03--d04. Because the LES domains receive comparatively smooth inflow from their coarser parents, they also require sufficient temporal spin-up and spatial fetch for resolved eddies to develop and for the resolved--SGS energy partition to equilibrate. Spin-up is therefore selected according to the dynamical role of each domain. Domain d01 primarily inherits the HRRR initial and boundary conditions and requires adjustment mainly for physics initialization and the model vertical grid. Domain d02 must additionally develop a dynamically consistent boundary-layer structure, including realistic shear and stratification. In d03 and d04, the spin-up period allows resolved turbulent motions to form before the fields are evaluated. Each simulation includes a total spin-up period of 12 hours before the analysis window. Resolving the innermost domain at $40~\mathrm{m}$ makes the simulations computationally demanding. On the high-performance computing cluster used for this study, a 24-hour simulation required approximately 12--18 hours of wall-clock time on 12--20 compute nodes, depending on resource availability. With the multiscale configuration established, the next section describes the SGS and SFS formulations that govern the unresolved contribution to turbulence in the LES domains.

\section{On the SGS and SFS for LES domain in WRF}

The WRF literature uses both \emph{subgrid-scale} (SGS) and
\emph{subfilter-scale} (SFS) often interchangeably,
but they need not to be identical in a discrete model.  ``Subgrid''
strictly refers to scales smaller than the mesh cutoff.  ``Subfilter''
can include both those scales and nominally resolvable scales that are
strongly attenuated by the numerical operators. The classic WRF--LES implementation discussed in \cite{Mirocha2010}
uses the mesh and numerical discretization as an implicit filter.
Consequently, ``SFS stress'' is the broader and more precise term,
while WRF documentation also uses ``sub-grid turbulent stress.''  Here, SGS is used for the modeled unresolved turbulence and SFS
when emphasizing the numerical filtering interpretation. A domain becomes LES-like through a consistent collection of
settings, grid choices, and boundary forcing.  Merely changing
\code{sfs\_opt} does not convert a mesoscale domain to LES in WRF. For the WRF ARW dynamical core, the relevant options that play distinct
roles are listed below. Various options of these parameters and their combinations are explained below and summarized in Table \ref{tab:matrix options}. 

\noindent $\bullet$ {\code{bl\_pbl\_physics}: one-dimensional PBL parameterization, if any}. A conventional mesoscale PBL scheme parameterizes nearly all turbulent
vertical transport in a one-dimensional column.  In a conventional
WRF LES domain it is turned off:
\code{bl\_pbl\_physics = 0} so that resolved three-dimensional eddies and a three-dimensional SGS
closure carry the turbulent fluxes. A parent mesoscale domain may retain a PBL scheme.  For example, for 4 nested domain with
\code{bl\_pbl\_physics = 1, 0, 0, 0,} we will have YSU on d01 and no PBL on d02--d04.

\noindent $\bullet$ {\code{diff\_opt}: how the three-dimensional diffusion/stress divergence is evaluated}. For
\code{diff\_opt=1}, the diffusion derivatives are evaluated along terrain-following coordinate
surfaces and for \code{diff\_opt=2}, 
full physical-space stress-form diffusion, including metric terms
needed to calculate gradients more consistently over sloping terrain.
This is the required pathway for nonlinear backscattering and anisotropy (NBA),  and is normally the relevant
choice for real-data WRF--LES.

\noindent $\bullet$ {\code{km\_opt}: how eddy diffusivities or the base SGS closure are obtained}. \code{km\_opt=2} is the three-dimensional, prognostic SGS-TKE closure.  WRF advances an SGS-TKE variable and derives eddy diffusivities from it. \code{km\_opt=3} is the three-dimensional, diagnostic deformation closure of Smagorinsky type. The mixing coefficient is diagnosed from resolved deformation and
stability. \code{km\_opt=4} is the two-dimensional horizontal deformation mixing.  Vertical mixing is
supplied by an active PBL scheme.  This is appropriate for a mesoscale
parent with a PBL scheme, not for a conventional LES domain.

\noindent $\bullet$ {\code{sfs\_opt}: whether the momentum SGS stress is replaced by an NBA tensor}.


\subsection{Eddy-viscosity closures}

LES resolves the energetic turbulent motions that are larger than the
effective filter width and models the influence of motions below that
width. In general, for a velocity component \(u_i\), we can write $u_i = \widetilde{u}_i + u_i'$, where \(\widetilde{u}_i\) is the resolved, filtered velocity and
\(u_i'\) denotes unresolved motion.  Filtering the nonlinear
advection term produces a stress tensor $\tau_{ij}
=
\widetilde{u_i u_j}
-
\widetilde{u}_i\,\widetilde{u}_j $.
The filtered momentum equation contains the divergence of this tensor:
\begin{equation}
\frac{\partial \widetilde{u}_i}{\partial t}
+
\widetilde{u}_j
\frac{\partial \widetilde{u}_i}{\partial x_j}
=
-\frac{1}{\rho_0}
\frac{\partial \widetilde{p}}{\partial x_i}
+\mathcal{F}_i
-
\frac{\partial \tau_{ij}}{\partial x_j},
\label{eq:filtered-momentum}
\end{equation}
where \(\mathcal{F}_i\) denotes buoyancy, Coriolis, and other resolved
forcings.  The exact \(\tau_{ij}\) cannot be computed from the
resolved velocity alone, so it must be closed. The isotropic part of \(\tau_{ij}\) is often absorbed into a modified
pressure.  The closure therefore focuses mainly on the deviatoric
part,
\begin{equation}
\tau_{ij}^{d}
=
\tau_{ij}
-\frac{1}{3}\tau_{kk}\delta_{ij}.
\end{equation}
Both the standard Smagorinsky and standard 1.5-order TKE closures use
a Boussinesq-type relation for the deviatoric momentum stress:
\begin{equation}
\tau_{ij}^{d}
=
-2K_m S_{ij},
\label{eq:linear-eddy-viscosity}
\qquad 
S_{ij}
=
\frac{1}{2}
\left(
\frac{\partial \widetilde{u}_i}{\partial x_j}
+
\frac{\partial \widetilde{u}_j}{\partial x_i}
\right),
\end{equation}
where $S_{ij}$ is the resolved strain-rate tensor and \(K_m\ge 0\) is an eddy
viscosity. The SGS transfer of resolved kinetic energy is commonly written in the form
$
\Pi
=
-\tau_{ij}^{d}S_{ij}
$, which by substituting $\tau_{ij}^{d}$ gives
\begin{equation}
\label{eq:pi}
\Pi = 2K_m S_{ij}S_{ij}\ge 0.
\end{equation}
With this sign convention, \(\Pi>0\) is forward transfer from resolved
motion to unresolved motion.  A positive linear eddy viscosity is
therefore purely dissipative at each point and instant.  This property
is robust numerically, but it prevents local reverse transfer and
forces the stress tensor to align with the resolved strain tensor. The namelist option \code{km\_opt} selects the basic turbulence
closure used to determine \(K_m\).


In schematic form, a Smagorinsky closure  (\code{km\_opt=3}) diagnoses
\begin{equation}
K_m
=
(C_s\Delta)^2
\left(2S_{ij}S_{ij}\right)^{1/2}
f_{\mathrm{stab}},
\label{eq:smag-k}
\end{equation}
where \(C_s\) is a model coefficient, \(\Delta\) is a filter-width
measure, and \(f_{\mathrm{stab}}\) represents stability effects.  The
closure is called diagnostic because \(K_m\) is obtained directly from
the current resolved gradients; no separate SGS-energy equation is
required. Its main advantages are simplicity and low cost.  Its principal
limitations are that the coefficient is not universally optimal, the
stress is constrained to be aligned with strain, normal-stress
anisotropy is not represented by the deviatoric closure, and local
backscatter is absent. 

The 1.5-order closure (\code{km\_opt=2}) advances a prognostic SGS TKE,
$e_{\sgs}
=
\frac{1}{2}
\widetilde{u_i'u_i'}$. 
A schematic SGS-TKE budget is
\begin{equation}
\frac{D e_{\sgs}}{Dt}
=
P_s + P_b
-
\varepsilon
+
\frac{\partial}{\partial x_j}
\left(
K_e\frac{\partial e_{\sgs}}{\partial x_j}
\right)
+\mathcal{T},
\label{eq:tke-budget}
\end{equation}
where \(P_s\) is shear production, \(P_b\) is buoyancy production or
destruction, \(\varepsilon\) is dissipation, and \(\mathcal{T}\)
collects additional modeled transport terms.  The eddy viscosity and
dissipation are commonly represented schematically as
$K_m \sim C_k\,\ell\,\sqrt{e_{\sgs}}$ and $
\varepsilon \sim C_\varepsilon
\frac{e_{\sgs}^{3/2}}{\ell}$ with mixing length \(\ell\). Here, ``1.5 order'' means that the first moment (mean/resolved state) and a
second-order quantity (TKE) are predicted, while higher-order
transport and dissipation terms are closed diagnostically.  The
prognostic TKE gives the closure a memory and allows advection of SGS
energy, but the ordinary momentum stress still uses the linear
stress--strain relation in \cref{eq:linear-eddy-viscosity} when
\code{sfs\_opt=0}.

\subsection{Nonlinear Backscatter and Anisotropy in WRF}

Nonlinear backscatter and anisotropy (NBA) identifies the two physical capabilities added to the ordinary linear
eddy-viscosity stress. \textbf{Nonlinear:} The modeled stress includes terms that are quadratic in the resolved
velocity-gradient tensors, rather than only a scalar eddy viscosity \(K_m\) multiplied by
 \(S_{ij}\).  Consequently, the tensor orientation is not forced
to be parallel to the strain tensor. \textbf{Backscatter:} The average three-dimensional turbulence cascade is predominantly
forward: energy moves from large resolved eddies toward progressively
smaller scales, where it is dissipated.  Locally and intermittently,
however, the transfer can reverse.  With the definition in
\cref{eq:pi}, backscatter occurs where
$\Pi < 0$. A strictly positive linear eddy viscosity cannot produce this local
state while the nonlinear NBA terms can. Backscatter should not be interpreted as arbitrary creation of energy.
It is a modeled local transfer from unresolved to resolved motions.
A viable SGS model must still maintain physically reasonable domain
and time-mean energy behavior and numerical stability. \textbf{Anisotropy:} Near the surface, in shear layers, under stratification, and over
complex terrain, unresolved turbulence is not generally isotropic.
The normal stresses may differ: $\tau_{11}\ne\tau_{22}\ne\tau_{33}$, and the stress and strain tensors need not share eigenvectors.  A
simple scalar eddy viscosity cannot represent this full tensor
behavior.  NBA adds traceless quadratic strain terms and
strain--rotation coupling terms that allow non-alignment and
anisotropic normal stresses.


To display the equations in the convention used in the WRF NBA
implementation, let's define deformation and rotation tensors
without the usual factor \(1/2\) as:
\begin{align}
S^{\star}_{ij}
=
\frac{\partial \widetilde{u}_i}{\partial x_j}
+
\frac{\partial \widetilde{u}_j}{\partial x_i},
\qquad
R^{\star}_{ij}
=
\frac{\partial \widetilde{u}_i}{\partial x_j}
-
\frac{\partial \widetilde{u}_j}{\partial x_i}.
\end{align}
Thus \(S^{\star}_{ij}=2S_{ij}\).The effective filter width is
$\Delta=(\Delta x\,\Delta y\,\Delta z)^{1/3}$.
The nonlinear part contains two important tensors:
\begin{align}
A_{ij}
=
S^{\star}_{ik}S^{\star}_{kj}
-\frac{1}{3}
S^{\star}_{mn}S^{\star}_{mn}\delta_{ij},
\qquad
B_{ij}
=
S^{\star}_{ik}R^{\star}_{kj}
-
R^{\star}_{ik}S^{\star}_{kj}.
\end{align}
The first is the traceless quadratic strain contribution; the second
is a strain--rotation coupling.  Together they permit stress
anisotropy and stress--strain misalignment. Factors of two and the sign of the rotation term depend on the tensor
convention used in a derivation or source code. The WRF options for SFS and the corresponding  equations below
follow the compact convention in the Mirocha et al.\ WRF implementation report \cite{Mirocha2010}.  





$\bullet$ With \code{sfs\_opt = 1} (NBA1), WRF computes the momentum SGS stress diagnostically from the
instantaneous resolved deformation and rotation. The current WRF
documentation requires
\code{diff\_opt = 2, 
km\_opt   = 2 or 3}. 
The word \emph{diagnostic} means that the NBA momentum stress does not
require its own prognostic stress equation or prognostic SGS-TKE
amplitude.  Its magnitude is determined from the local velocity
gradients and \(\Delta\). In the implementation-report convention, the compact NBA1 equation is
\begin{equation}
M_{ij}^{\mathrm{NBA1}}
=
-(C_s\Delta)^2
\Bigg[
2
\left(
2S^{\star}_{mn}S^{\star}_{mn}
\right)^{1/2}
S^{\star}_{ij}
+
C_1
\left(
S^{\star}_{ik}S^{\star}_{kj}
-\frac{1}{3}
S^{\star}_{mn}S^{\star}_{mn}\delta_{ij}
\right)
+
C_2
\left(
S^{\star}_{ik}R^{\star}_{kj}
-
R^{\star}_{ik}S^{\star}_{kj}
\right)
\Bigg],
\label{eq:nba1}
\end{equation}
where the first term inside paranteses is a Smagorinsky-like dissipative contribution, and the
second and third terms are the nonlinear anisotropy and
backscatter-capable terms. We note that in WRF, \code{sfs\_opt} changes the SGS \emph{momentum stress} and the potential temperature, water vapor, hydrometeors, and other scalars
still need SGS diffusivities.  The official namelist description
therefore states that NBA1 uses \code{km\_opt=2} or 3 ``for scalars.'' The combinations have different scalar physics and thus, \code{km\_opt} is still required and not redundant when NBA1 is active; see Table \ref{tab:matrix options}
for the summary of options. 


\begin{longtable}{>{\ttfamily}p{1.2cm}>{\ttfamily}p{1.2cm}>{\ttfamily}p{1.35cm}p{4.4cm}p{4.2cm}}
\caption{Interpretation of the principal \code{sfs\_opt} combinations.
NBA requires \code{diff\_opt=2}.}
\label{tab:matrix options}\\
\toprule
sfs\_opt & km\_opt & diff\_opt & Momentum SGS treatment & Scalar SGS treatment\\
\midrule
\endfirsthead
\toprule
sfs\_opt & km\_opt & diff\_opt & Momentum SGS treatment & Scalar SGS treatment\\
\midrule
\endhead
0 & 2 & 2 &
Linear eddy-viscosity stress with prognostic 1.5-order SGS TKE &
Diffusivity based on prognostic SGS TKE\\

0 & 3 & 2 &
Linear Smagorinsky/deformation stress &
Smagorinsky/deformation diffusivity\\

1 & 2 & 2 &
NBA1 diagnostic nonlinear stress from deformation and rotation &
Diffusivity based on prognostic SGS TKE\\

1 & 3 & 2 &
NBA1 diagnostic nonlinear stress from deformation and rotation &
Smagorinsky/deformation diffusivity\\

2 & 2 & 2 &
NBA2 nonlinear stress whose leading amplitude is based on prognostic SGS TKE &
Diffusivity based on prognostic SGS TKE\\

2 & 3 & 2 &
Not the documented NBA2 combination; NBA2 requires prognostic SGS TKE &
Not recommended/documented for NBA2\\

1 or 2 & any & 1 &
Not the documented NBA pathway; NBA requires full stress-form diffusion &
---\\
\bottomrule
\end{longtable}

$\bullet$ With \code{sfs\_opt = 2} (NBA2), the momentum stress uses the prognostic SGS TKE to set the amplitude
of its leading stress contribution.  The required combination is \code{diff\_opt = 2, 
km\_opt   = 2} because \code{km\_opt=2} supplies \(e_{\sgs}\). The implementation-report form for the compact NBA2 equation is
\begin{equation*}
M_{ij}^{\mathrm{NBA2}}
=
-C_e\Delta
\Bigg[
2\sqrt{e_{\sgs}}\,S^{\star}_{ij}
+
\left(\frac{27}{8\pi}\right)^{1/3}
C_s^{2/3}\Delta
\Bigg\{
C_1
\left(
S^{\star}_{ik}S^{\star}_{kj}
-\frac{1}{3}
S^{\star}_{mn}S^{\star}_{mn}\delta_{ij}
\right)
+
C_2
\left(
S^{\star}_{ik}R^{\star}_{kj}
-
R^{\star}_{ik}S^{\star}_{kj}
\right)
\Bigg\}
\Bigg],
\end{equation*}
where the nonlinear tensor structure is similar to NBA1, but, the decisive
difference is the velocity scale. NBA1 obtains its stress magnitude entirely from resolved deformation and grid scale while NBA2 uses the prognosed SGS energy \(\sqrt{e_{\sgs}}\) in the leading term and therefore carries SGS-energy history and transport through the TKE equation. The Mirocha et al.\ implementation report \cite{Mirocha2010} gives one coefficient set
based on the Kosovi\'c model \cite{Kosovic1997}, including a backscatter coefficient.

\begin{wrapfigure}{r}{0.4\textwidth}
\centering
\vspace{-0.3in}
\includegraphics[width=0.8\linewidth,trim={1.45cm 1.1cm 0.5cm 0.55cm},clip]{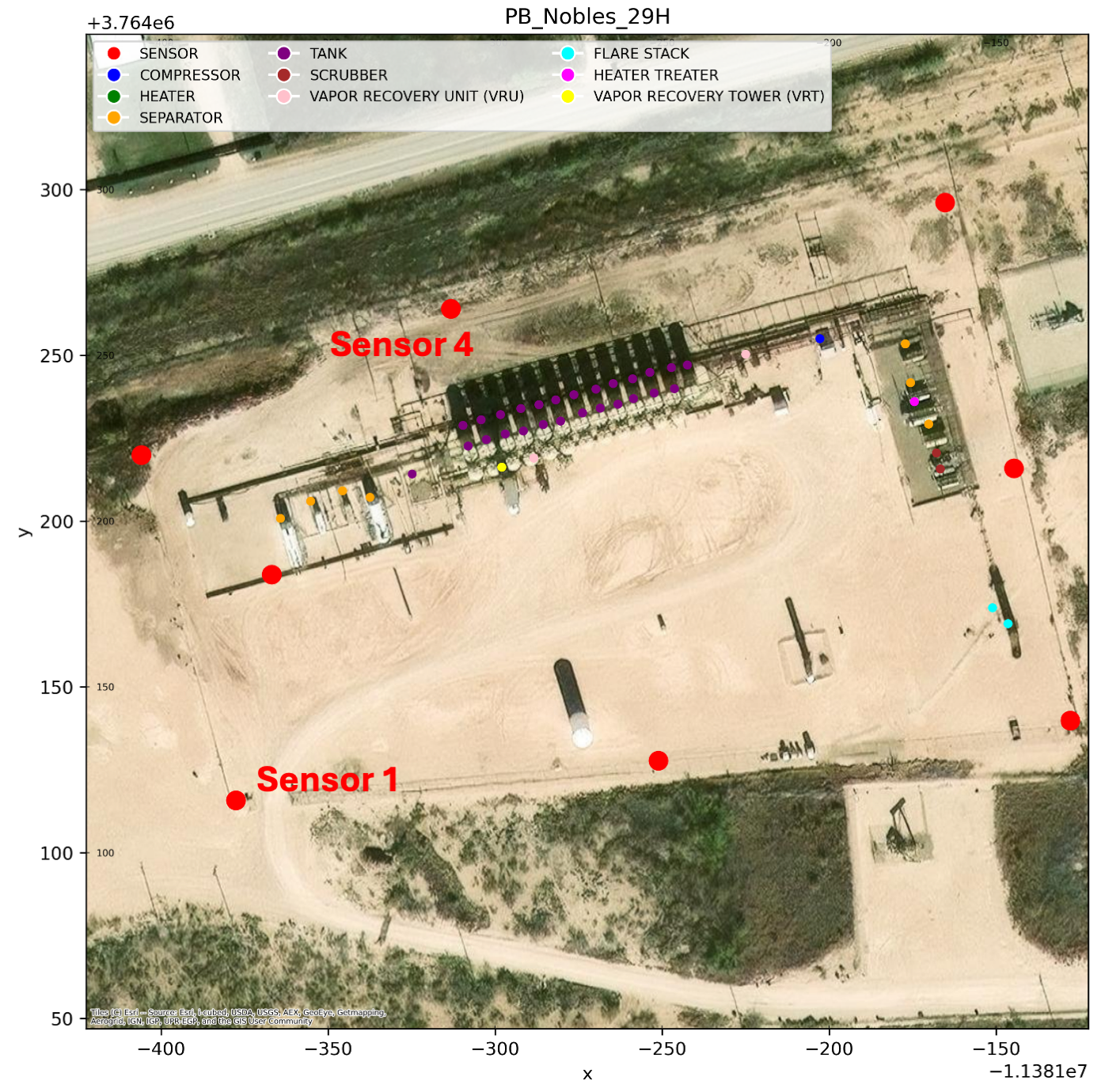}
\vspace{-0.05 in}
\caption{Schematic of location of the two wind sensors on the PB-NOBLES-29H site. The locations of other features on the site are also noted, including 8.3 meter tall tanks in the cluster of dark purple points and other sensors (red) that measure methane concentration only.}
\label{fig:pbn_layout_wD4}
\vspace{-0.5 in}
\end{wrapfigure}
We note that the exact coefficient definitions should be checked against the source
for the WRF version being run because implementation conventions and
code organization may differ across versions. The option-by-option physics matrix and brief descriptions are provided in Table \ref{tab:matrix options}.

\section{Field Sensor Network and Vertical Considerations in Boundary Layer}
\label{sec:field sensor}
\begin{figure}[h]
\centering
\includegraphics[width=0.8\linewidth]{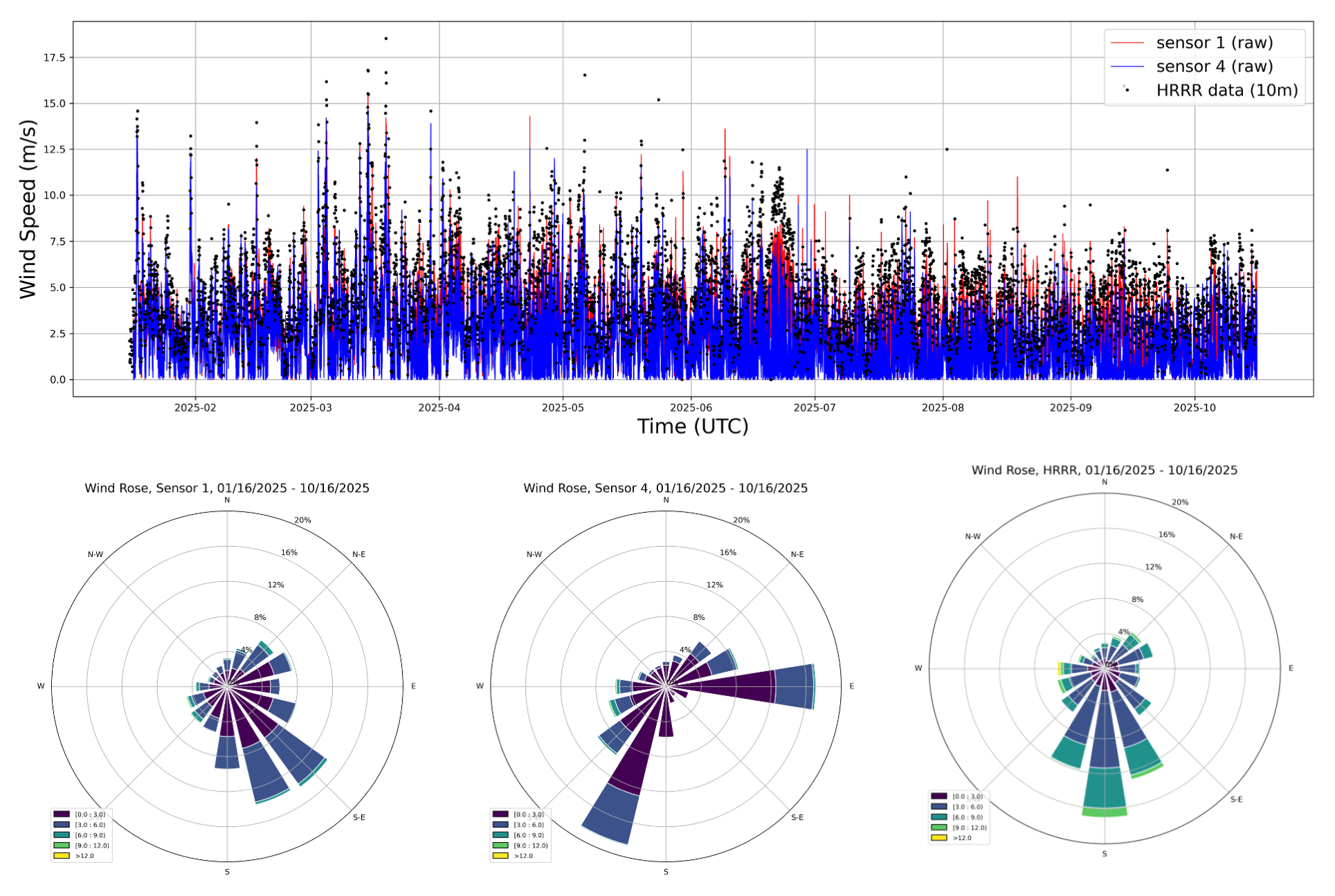}
\caption{Data collected every minute from two wind sensors located on the O\&G site plotting with HRRR data collected hourly. Wind roses show the direction the wind blows from as well as the wind speed.}
\label{fig:sensors}
\end{figure}

The field sensor network at the 
considered O\&G site comprises two anemometer stations, denoted Sensor~1 and Sensor~4, each mounted $1.5~\mathrm{m}$ above ground level and recording wind speed and direction at one-minute intervals. \Cref{fig:pbn_layout_wD4} shows the schematic location of the sensor with respect to other equipments in the field. To characterize the prevailing wind regimes at the site, we analyzed nearly a full year of measurements (16 January--16 October 2025) alongside the collocated HRRR wind field sampled hourly at $10~\mathrm{m}$. \Cref{fig:sensors} summarizes this record: the upper panel shows the raw wind-speed time series from both sensors together with the HRRR data, while the lower panels present the corresponding wind roses. The time series shows that the site experiences frequent moderate winds, typically in the $2$--$6~\mathrm{m/s}$ range, punctuated by stronger events exceeding $15~\mathrm{m/s}$ that occur most often during the winter and spring months. The wind roses, however, reveal a systematic discrepancy between the point measurements and the coarse HRRR field: both sensors record prevailing winds from the south and southeast, with a secondary easterly component, whereas HRRR indicates a predominantly northerly flow at consistently higher speeds. This mismatch in both direction and magnitude underscores the limitations of using coarse-resolution products directly at the facility scale and motivates the dynamical downscaling and observational nudging described in \Cref{sec:nudging}. Two additional features of the observations warrant note. First, local structures such as storage tanks are present on the site but are not represented in the WRF terrain, so their sheltering and flow-distortion effects on the near-surface wind are not captured by the model. Second, the Sensor~4 wind rose exhibits a pronounced directional gap, with few winds recorded from the western sectors; this gap is consistent with flow obstruction by nearby site structures and illustrates how strongly an individual point anemometer can be influenced by its immediate surroundings.


\begin{figure}[h]
\centering
\includegraphics[width=0.6\linewidth]{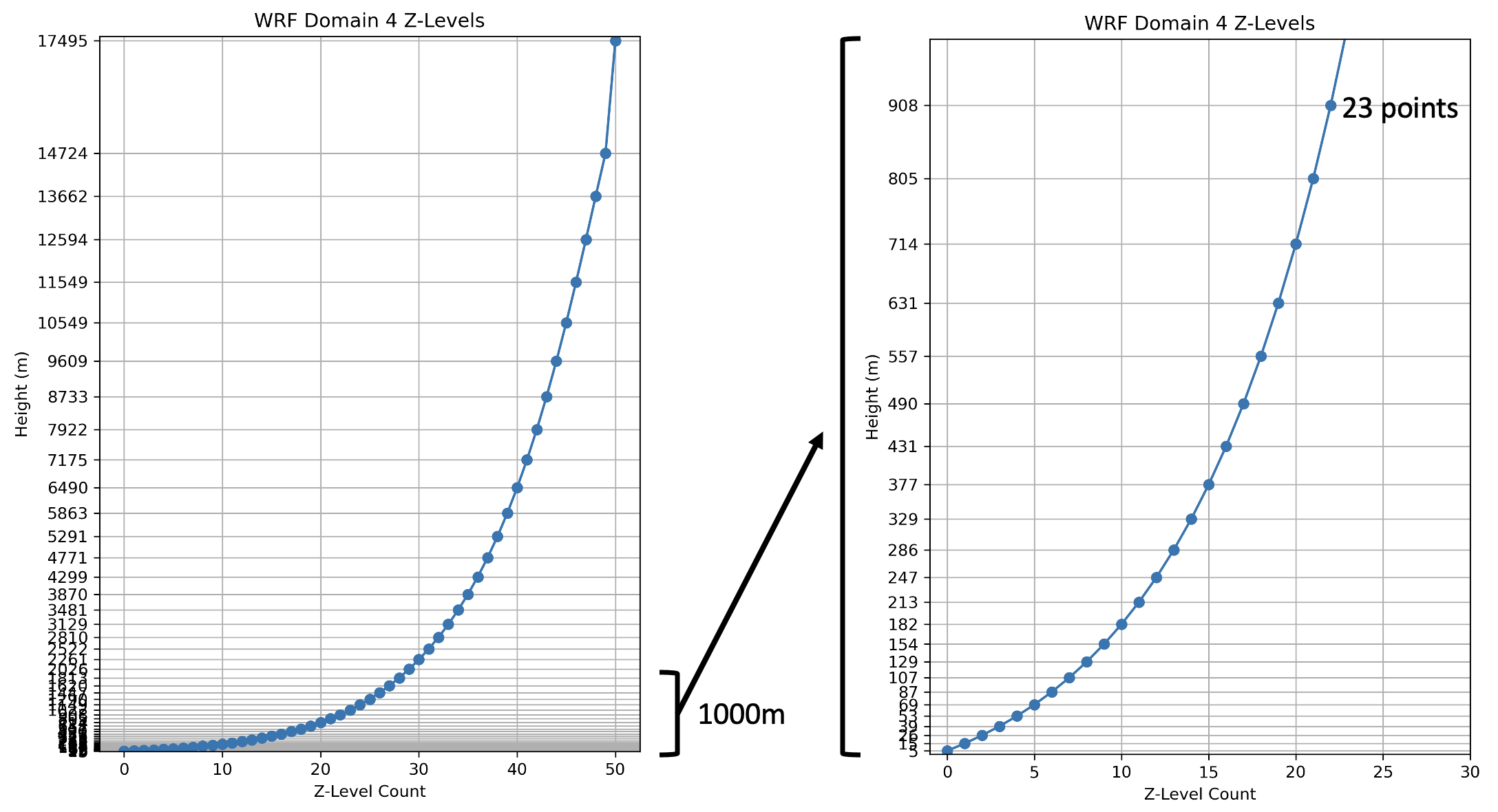}
\caption{Distribution of points along z direction in height in the WRF simulation setup. There are 23 points in the first 1000 m height. }
\label{fig:zlevel}
\end{figure}

The Texas Permian Basin region exhibits pronounced diurnal variability in boundary-layer structure. Under strong daytime surface heating and typically dry atmospheric conditions, the convective boundary layer (CBL) over Midland, Texas can deepen to approximately 1.5–3 km above ground level (AGL). At night, radiative cooling of the surface leads to the formation of a shallow, stable boundary layer (SBL), often confined to roughly 100–500 m AGL. Accurately resolving this strong diurnal evolution is essential for simulating near-surface wind shear, turbulent kinetic energy (TKE), and plume dispersion relevant to oil and gas facilities. To capture this variability, our WRF configuration employs 54 vertical levels (\code{e\_vert = 54}) extending from the surface to a model top of 5000 Pa (\code{p\_top\_requested = 5000 Pa}), corresponding to approximately 20 km altitude. The distribution of points along z direction in height in the WRF simulation setup is shown in \Cref{fig:zlevel}. We can see that we use 23 points in the first 1000 m height. The model uses a hybrid sigma–pressure vertical coordinate (\code{hybrid\_opt = 2}), which transitions from terrain-following coordinates near the surface to pressure-based coordinates aloft. This hybrid formulation reduces numerical distortions over complex terrain while preserving fine vertical resolution within the boundary layer, where steep gradients in wind speed, temperature, and moisture occur.

In the outer mesoscale domains (d1–d2), boundary-layer turbulence is parameterized using the YSU PBL scheme, which diagnoses boundary-layer depth and vertical mixing based on bulk Richardson number and entrainment considerations. In the inner LES domains (d3–d4), the PBL scheme is disabled and turbulence is represented by resolved eddies and a 1.5-order TKE-based subgrid-scale closure (km\_opt = 2, sfs\_opt = 2). Consequently, adequate vertical resolution near the surface is critical for allowing shear production of turbulence and realistic vertical transport of momentum and scalars within the first few hundred meters, particularly during stable nighttime conditions when vertical gradients are strongest. WRF maps physical height to terrain-following eta ($\eta$) coordinates according to
\[
\eta = \frac{p - p_{top}}{p_s - p_{top}}
\]
where $p$ is pressure at a given model level, $p_s$ is surface pressure, and $p_{top}$ is the model top pressure. Because eta surfaces follow terrain near the ground, sensor measurement heights (e.g., 2 m, 10 m, or tower levels) must be converted to geometric height using model geopotential fields and then matched to the corresponding eta levels for diagnostics and data assimilation. This translation is particularly important in the LES domains, where resolved turbulence and strong vertical shear within the lowest 100–300 m can produce substantial vertical variability over short distances. The combination of 54 vertical levels, a high model top, and hybrid terrain-following coordinates ensures that both shallow nocturnal boundary layers and deep daytime convective layers are represented within the same simulation framework, while maintaining vertical resolution sufficient for turbulence-resolving dynamics in the innermost domain.

\section{Observations Nudging and Two-Way Nesting in WRF Model} \label{sec:nudging}


While the WRF model is initialized and driven by boundary conditions derived from HRRR data, the resulting high-resolution wind fields can still drift from the local conditions observed on site. Nudging introduces a relaxation term into the governing equations to constrain the modeled state using externally supplied information. Analysis (grid) nudging relaxes selected model variables toward a gridded analysis, whereas observational nudging relaxes them toward individual observations within prescribed spatial and temporal influence windows. These approaches can serve complementary roles in a multiscale assimilation strategy, with large-scale analyses constraining the outer flow and local observations constraining finer-scale features \cite{StaufferSeaman1994}. The distinction is important here because assimilating the site anemometers is a different task from retaining consistency with the regional HRRR forcing.

To improve our simulation results, we incorporate \emph{observational nudging} using the wind-sensor measurements collected from two anemometers at the site. We apply this nudging in the innermost domain (d04), which has a grid spacing of $40~\mathrm{m}$ and is run in large-eddy simulation (LES) mode. The observational nudging modifies the model tendencies by adding a relaxation term:
\[
\frac{d\phi}{dt} = F_{\text{model}} + G \, W(\mathbf{x}, t)\left(\phi_{\text{obs}} - \phi_{\text{model}}\right),
\]
where $\phi$ is a prognostic variable (here, the horizontal wind components), $F_{\text{model}}$ denotes the physical and dynamical tendencies computed by the model, $G$ is the nudging coefficient (an inverse relaxation time scale), and $W(\mathbf{x}, t)$ is a weighting function that decays with spatial and temporal distance from each observation.

Previous studies have found that nudging is most effective at larger scales, and it is most commonly applied above the planetary boundary layer (PBL), where it can steer the large-scale flow without disturbing resolved turbulence. Its use at finer, turbulence-resolving scales is less established. Two case studies are particularly relevant to the present configuration. In the first, Doppler-lidar observations were assimilated through observational nudging in a gray-zone WRF simulation over Osaka, Japan \cite{NayakKanda2023}. That study used a single domain with $500~\mathrm{m}$ grid spacing on a $100\times100$ grid (a $50~\mathrm{km}\times50~\mathrm{km}$ domain) and $40$ vertical levels, with observations at $43$, $128$, and $214~\mathrm{m}$. The authors obtained reasonable results, although their configuration is best characterized as very-high-resolution mesoscale (gray-zone) modeling rather than LES. Within fully developed LES domains, nudging is generally discouraged because it can damp the resolved turbulent structures that the LES is intended to capture. A second study, by Liu et al.\ \cite{Liu2020}, evaluated flow over complex terrain with WRF--LES at $37~\mathrm{m}$ grid spacing, close to our own $40~\mathrm{m}$ d04 domain, and reported that the model could reasonably capture large-scale events and microscale circulation characteristics. That study assessed sensitivity to boundary-layer and subgrid-scale turbulence treatments rather than near-surface observational nudging; it therefore supports the use of WRF--LES at comparable grid spacing, not the effectiveness of nudging within an LES domain. Here, we investigate wind-only observational nudging in the innermost domain and assess its effect on agreement with the sensor data in \Cref{sec:experiments}.

\subsection{Implementation of Observational Nudging}

We employ four-dimensional data assimilation (FDDA) in the form of observation nudging within WRF. Nudging is applied exclusively in the innermost LES domain (d04; 40 m grid spacing) (\code{obs\_nudge\_opt = 1} for d04), and only to the wind components (\code{obs\_nudge\_wind = 1}), while temperature and moisture are left unconstrained. The nudging tendency relaxes the model wind toward observations with a coefficient of 0.12 (\code{obs\_coef\_wind = 0.12}), introducing a gradual correction term in the prognostic momentum equations rather than imposing a hard constraint. Observations are assimilated within a limited spatial radius of influence (\code{obs\_rinxy} in the inner domains) and a short temporal window (\code{obs\_twindo}), ensuring that corrections remain localized in space and time.

\begin{figure}[t]
\centering
\includegraphics[width=1\linewidth]{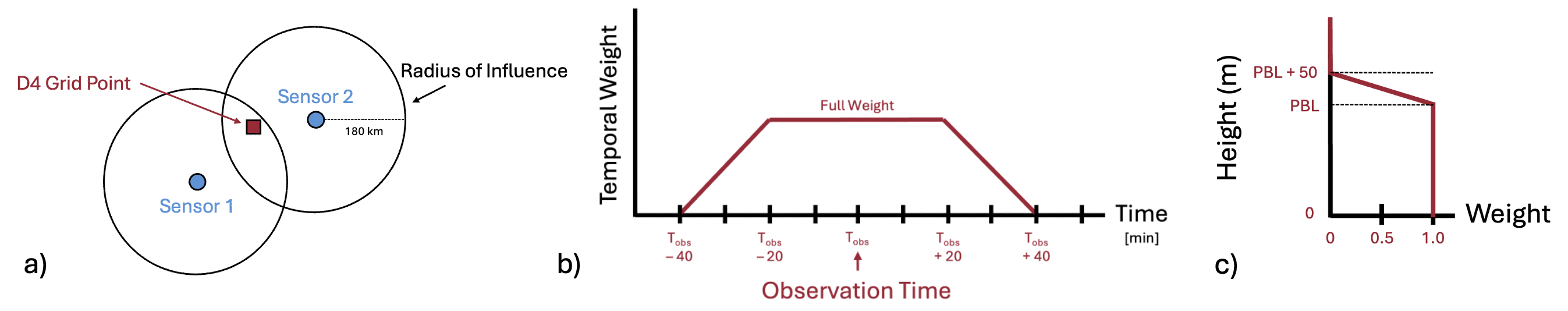}
\caption{Nudging parameters as listed in Table \ref{tab:wrf_nudging}, a) Horizontal Radius of Influence, b) Temporal Window for Observation influence, c) Vertical Radius of Influence}
\label{fig:nudgingParams}
\end{figure}

\begin{table*}[t]
\centering
\caption{WRF nudging configuration.}
\label{tab:wrf_nudging}
\begin{threeparttable}
\small
\setlength{\tabcolsep}{6pt}
\renewcommand{\arraystretch}{1.15}
\begin{tabularx}{\textwidth}{
@{}
>{\raggedright\arraybackslash}X
>{\centering\arraybackslash}p{0.15\textwidth}
>{\centering\arraybackslash}p{0.15\textwidth}
>{\centering\arraybackslash}p{0.15\textwidth}
>{\centering\arraybackslash}p{0.15\textwidth}
@{}
}
\toprule
\textbf{Nudging}
& \textbf{d01}
& \textbf{d02}
& \textbf{d03}
& \textbf{d04} \\
\midrule
Grid-analysis nudging
& Off
& Off
& Off
& Off \\
Observation nudging
& Off
& Off
& Off
& Wind only \\
Wind-observation nudging coefficient
& --
& --
& --
& 0.12 s$^{-1}$ \\
Horizontal radius of influence (km)
& --
& --
& --
& 180 
\\
Temporal window for observation influence (min)
& --
& --
& --
& 40 
\\
\bottomrule
\end{tabularx}
\end{threeparttable}
\end{table*}

This selective wind-only nudging strategy is designed to reduce large-scale directional and speed biases inherited from the parent domains or driving HRRR fields, while allowing the LES to freely generate turbulence through resolved shear production and buoyancy. Because the PBL scheme is disabled in the LES domains (\code{bl\_pbl\_physics = 0}), turbulence is represented through resolved eddies and the 1.5-order TKE-based subgrid-scale closure (\code{km\_opt = 2}, \code{sfs\_opt = 2}). The nudging term therefore primarily adjusts the mean flow and low-wavenumber components of the wind field, while small-scale fluctuations remain governed by the resolved dynamics and SGS dissipation. This approach balances two competing objectives: (i) maintaining realistic mean wind direction and magnitude at the site for accurate plume transport and source inversion, and (ii) preserving physically consistent resolved turbulent kinetic energy necessary for dispersion modeling. By confining nudging to the innermost domain and to wind variables only, we minimize artificial damping of turbulence and avoid suppressing scalar transport variability critical for emission detection applications.



WRF observational nudging requires the observations to be supplied in the Little-R format, which is subsequently converted into the \code{obsnud} binary format read at runtime. The raw sensor data (CSV format), containing wind speed $U$ and direction $\theta$, were converted to the $u$- and $v$-components following 
$u = -U \sin(\theta)$ and $v = -U \cos(\theta)$, where $\theta$ is the meteorological wind direction (in degrees clockwise from north), i.e. the direction the wind is blowing \textit{from}. 
%
%
%
%
Each observation influences the model grid points that lie within a horizontal radius of influence $R$ and time window $\Delta t$, and the relaxation time scale $\tau = 1/G$ sets how strongly the observation constrains those grid points. We conducted sensitivity experiments, varying $R$, $\Delta t$, and $G$, to characterize the influence of each parameter on the simulation and to select the configuration reported in \Cref{tab:wrf_nudging}.











%
\begin{figure}[t]
\centering


\begin{minipage}[t]{0.48\textwidth}
\centering

\resizebox{\linewidth}{!}{%
\begin{tikzpicture}[
  node distance=1.25cm,
  box/.style={
    draw,
    rounded corners,
    minimum width=2.4cm,
    minimum height=0.9cm,
    align=center,
    thick
  },
  down/.style={-{Latex[length=3mm]}, thick, black!65},
  obsarr/.style={-{Latex[length=3mm]}, very thick, obsc}
]

\node[box, fill=d01c] (d01) {d01 (\SI{3}{km})};
\node[box, fill=d02c, below=of d01] (d02) {d02 (\SI{1}{km})};
\node[box, fill=d03c, below=of d02] (d03) {d03 (\SI{200}{m})};
\node[box, fill=d04c, below=of d03] (d04)
  {d04 (\SI{40}{m})\\directly obs nudged};

\node[right=1cm of d04, text=obsc, font=\bfseries]
  (obs) {Wind observations};

\draw[down]
  (d01) -- node[right, font=\small]{parent forcing} (d02);

\draw[down]
  (d02) -- node[right, font=\small]{parent forcing} (d03);

\draw[down]
  (d03) -- node[right, font=\small]{parent forcing} (d04);

\draw[obsarr] (obs) -- (d04);

\node[
  left=0.6cm of d01,
  rotate=90,
  font=\normalsize
] {
  fine forcing  $\leftarrow$ coarse 
};

\end{tikzpicture}%
}
\end{minipage}
\hfill
\begin{minipage}[t]{0.48\textwidth}
\centering

\resizebox{\linewidth}{!}{%
\begin{tikzpicture}[
  node distance=1.35cm,
  box/.style={
    draw,
    rounded corners,
    minimum width=2.7cm,
    minimum height=0.95cm,
    align=center,
    thick
  },
  down/.style={-{Latex[length=3mm]}, thick, black!65},
  up/.style={-{Latex[length=3mm]}, very thick, feedc},
  obsarr/.style={-{Latex[length=3mm]}, very thick, obsc}
]

\node[box, fill=d01c] (d01) {d01 (\SI{3}{km})};
\node[box, fill=d02c, below=of d01] (d02) {d02 (\SI{1}{km})};
\node[box, fill=d03c, below=of d02] (d03) {d03 (\SI{200}{m})};
\node[box, fill=d04c, below=of d03] (d04)
  {d04 (\SI{40}{m})\\directly obs nudged};

\node[right=1cm of d04, text=obsc, font=\bfseries]
  (obs) {Wind observations};

\draw[down]
  ([xshift=-5pt]d01.south) --
  ([xshift=-5pt]d02.north);

\draw[up]
  ([xshift=5pt]d02.north) --
  ([xshift=5pt]d01.south);

\draw[down]
  ([xshift=-5pt]d02.south) --
  ([xshift=-5pt]d03.north);

\draw[up]
  ([xshift=5pt]d03.north) --
  ([xshift=5pt]d02.south);

\draw[down]
  ([xshift=-5pt]d03.south) --
  ([xshift=-5pt]d04.north);

\draw[up]
  ([xshift=5pt]d04.north) --
  ([xshift=5pt]d03.south);

\draw[obsarr] (obs) -- (d04);

\node[
  left=0.6cm of d01,
  rotate=90,
  font=\normalsize
] {
  fine forcing  $\leftarrow$ coarse 
};

\node[
  right=0.5cm of d03,
  rotate=90,
  text=feedc,
  font=\normalsize\bfseries
] {
  fine $\rightarrow$ coarse feedback
};

\end{tikzpicture}%
}

\end{minipage}
\caption{Left: one-way nest \code{feedback = 0}, the nudged d04 state does not
overwrite or update d03 through the nesting algorithm. Right: two-way nesting \code{feedback = 1}, the nudged d04 will update back the parents domains. 
}
\label{fig:1-2-way}
\end{figure}
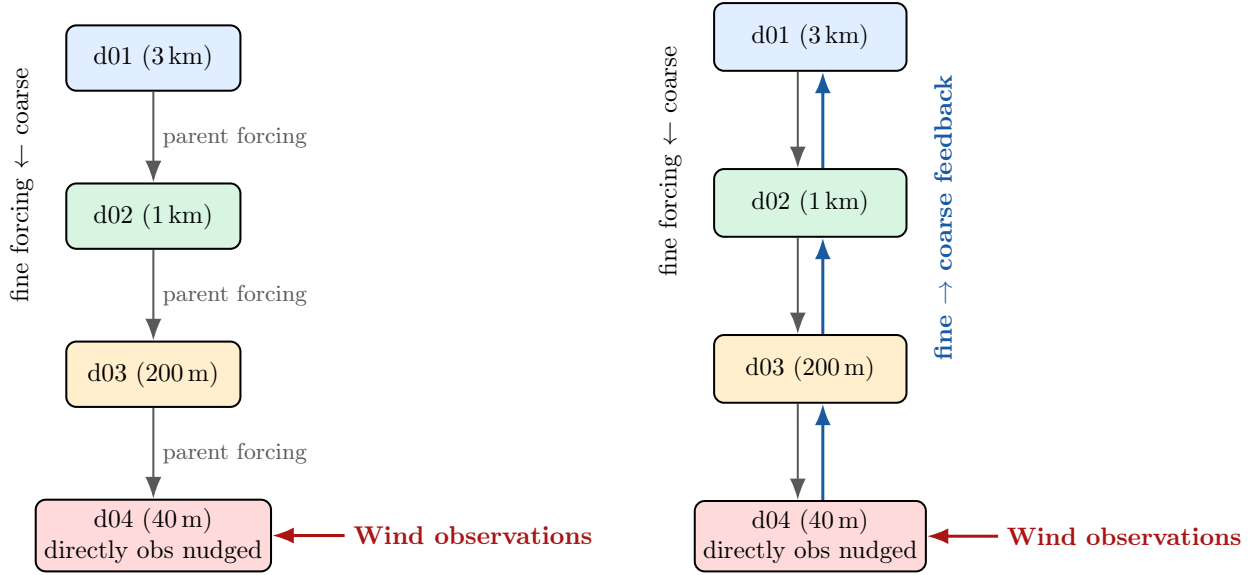

\subsection{Two-Way Nesting}
\label{sec:2wnest}
Moeng et al.\ \cite{Moeng2007} evaluated two-way nesting in idealized WRF LES-within-LES experiments, providing a basis for examining interactions between coarse and fine representations of boundary-layer turbulence. Their experiments address nested LES coupling rather than observational assimilation; in the present study, we examine how that coupling transmits a state modified by wind observations. Direct observational nudging is restricted to d04, and without nesting feedback its influence does not propagate back to the parent domains. Here, we explain the two-way nesting as a mechanism in WRF setup that allows the fine-grid atmospheric state resulting from wind observations assimilated only on the \SI{40}{m} d04 LES domain to propagate back into d03, d02, and d01. With one-way nesting (\code{feedback = 0}), information travels from parent to child through initialization and lateral boundary forcing. The child does not modify the parent. While with two-way nesting (\code{feedback = 1}), the parent still forces the child, but the child also feeds its calculated state back to the parent when the domains synchronize. The important consequence is that \texttt{feedback = 1} is a nesting setting for the telescoping hierarchy. It is not restricted to d04 $\rightarrow$ d03. Therefore the model permits information modified on d04 to influence d03, then the resulting d03 solution to influence d02, and then the resulting d02 solution to influence d01. \Cref{{fig:1-2-way}} schematically shows the hierarchy in the one- and two-way nesting setting with observational nudging in WRF.

\begin{wrapfigure}{r}{0.31\textwidth}
\centering
\vspace{-1.5em}
\begin{minipage}[t]{0.22\textwidth}
\centering
\resizebox{\linewidth}{!}{
\begin{tikzpicture}[
  node distance=0.5cm and 1.2cm,
  stage/.style={draw, rounded corners, align=center, minimum width=3.5cm, minimum height=0.85cm, thick},
  arr/.style={-{Latex[length=2mm]}, thick},
  farr/.style={-{Latex[length=2mm]}, very thick, feedc}
]
\node[stage, fill=red!8] (obs) {Station wind observation};
\node[stage, fill=d04c, below=of obs] (nudge) {d04 obs-nudging tendency\\changes d04 wind};
\node[stage, fill=d04c, below=of nudge] (evolve4) {d04 dynamics/physics evolve\\the nudged state};
\node[stage, fill=d03c, below=of evolve4] (feed43) {Feedback average/remap\\d04 $\rightarrow$ d03};
\node[stage, fill=d03c, below=of feed43] (evolve3) {d03 dynamics evolve\\the modified parent state};
\node[stage, fill=d02c, below=of evolve3] (feed32) {Feedback\\d03 $\rightarrow$ d02};
\node[stage, fill=d02c, below=of feed32] (evolve2) {d02 dynamics evolve\\the modified parent state};
\node[stage, fill=d01c, below=of evolve2] (feed21) {Feedback\\d02 $\rightarrow$ d01};

\draw[arr, obsc] (obs) -- (nudge);
\draw[arr] (nudge) -- (evolve4);
\draw[farr] (evolve4) -- (feed43);
\draw[arr] (feed43) -- (evolve3);
\draw[farr] (evolve3) -- (feed32);
\draw[arr] (feed32) -- (evolve2);
\draw[farr] (evolve2) -- (feed21);
\end{tikzpicture}
}
\end{minipage}
\vspace{-0.1 in}
\caption{Observation influence hierarchy pathway via a two-way nesting mechanism in WRF.}
\label{fig:2-way-pathway}
\vspace{-0.3 in}
\end{wrapfigure}
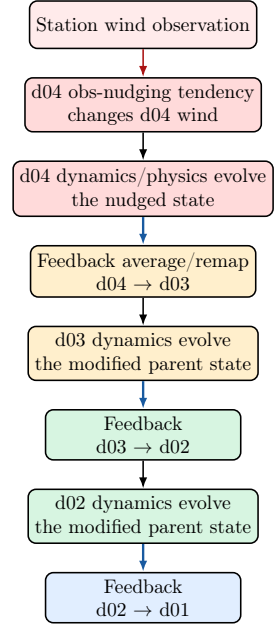
To better understand how this mechanism works in WRF, let's distinguish the \emph{nudging tendency} from the \emph{model state} and assume that d04 wind equation can be written as
\begin{equation}
\frac{\partial \mathbf{u}_{04}}{\partial t}
=
\mathcal{D}_{04}
+
\mathcal{P}_{04}
+
\mathcal{N}_{\mathrm{obs},04},
\end{equation}
where $\mathcal{D}_{04}$ represents resolved dynamics and numerical transport, $\mathcal{P}_{04}$ represents physics and subgrid-scale tendencies, and $\mathcal{N}_{\mathrm{obs},04}$ is the observational nudging tendency. During two-way nesting with \code{feedback = 1}, WRF feedback does not send a separate object called ``the observation correction'' to d03. Instead, after d04 has evolved under all of these tendencies, the resulting fine-grid state is spatially averaged/remapped back onto corresponding coarse-grid locations in the parent overlap region. For a generic parent state variable $X_{03}$, the operation can be represented schematically as
\begin{equation}
X_{03}^{\mathrm{new}}
\;\leftarrow\;
\mathcal{A}_{04\rightarrow03}\left(X_{04}^{\mathrm{evolved}}\right)
\qquad \text{within the d04-covered portion of d03},
\end{equation}
where $\mathcal{A}_{04\rightarrow03}$ denotes the WRF fine-to-coarse feedback averaging/remapping operation. Thus, d03 receives a state that already contains the combined effects of observation nudging plus d04 dynamics plus d04 physics plus d04 turbulence/subgrid response. For mass-grid quantities, WRF documentation describes coarse values being replaced using averages of fine-grid values at coincident locations; horizontal momentum uses corresponding fine-grid cell-face averaging. The exact operation is therefore a nested-grid state update, not a second application of the observational nudging operator on the parent. \Cref{{fig:2-way-pathway}} shows how the observation influence propagates through the hierarchy. The observation is directly assimilated only on d04 and d01--d03 are not given the d04 value of \code{obs\_rinxy} as their own nudging radius. The first direct fine-to-coarse modification occurs where d04 overlaps d03. After feedback, d03 dynamics can advect, mix, propagate, and otherwise redistribute the modified state beyond the immediate d04 footprint. The resulting d03 state can then influence d02 through the same two-way nesting mechanism, and similarly d02 can influence d01. Thus, the observational influence becomes progressively less like a direct station constraint and more like a dynamically processed model-state perturbation as it moves to coarser domains.

\begin{wrapfigure}{r}{0.35\textwidth}
\centering
\vspace{-5.5em}
\includegraphics[width=0.35\textwidth,trim={1.45cm 1cm 0.5cm 0.55cm},clip]{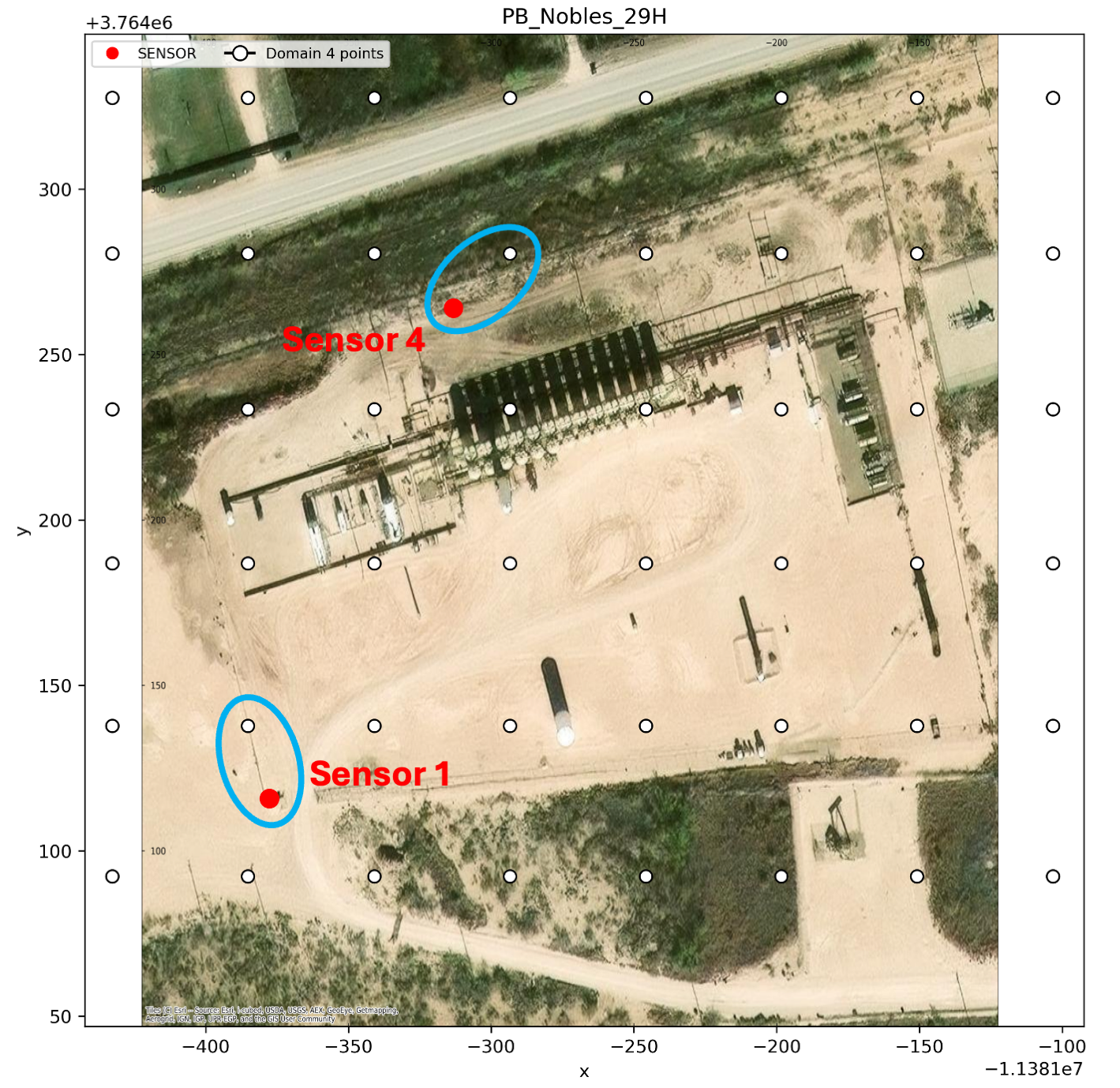}
\vspace{-0.05 in}
\caption{The raw sensor data were compared to the WRF simulation Domain 4 grid points closest to each sensor, as circled in blue.}
\label{fig:d4OverSite}
\vspace{-0.2 in}
\end{wrapfigure}
%
\section{Results and Validation} \label{sec:experiments}
This section evaluates the WRF-generated wind fields across multiple model configurations and meteorological periods by comparison with measurements from the two on-site anemometers. The cases include the baseline WRF simulation, simulations with observational nudging applied in the innermost domain (d04; $40~\mathrm{m}$ grid spacing), and sensitivity tests involving alternative subfilter-scale and nesting options. We examine both winter and summer periods---16 January at 18:00 UTC through 21 January at 18:00 UTC, and 1--5 July 2025---to determine whether the model behavior and the effect of nudging remain consistent under different wind regimes. Both Sensor~1 and Sensor~4 are supplied to the observational-nudging procedure and are subsequently used in the pointwise comparisons. For each case, the simulated fields are sampled at the d04 grid points nearest the two sensors, as illustrated in \Cref{fig:d4OverSite}, and are evaluated using wind-speed time series, power spectral density (PSD), spatial flow fields, and the quantitative metrics summarized in \Cref{tab:metrics_s1_d4,tab:metrics_s4_d4}. These comparisons assess agreement with the assimilated observations and the model's ability to represent the temporal variability and small-scale flow structure relevant to methane-plume transport; they should not be interpreted as an independent validation at the nudging locations. We begin with the direct time-series comparison and the effect of observational nudging and two-way nesting.


\subsection{Near-surface wind-speed response to two-way nesting without observational nudging}

\begin{figure}[h]
\centering
\includegraphics[width=1\linewidth]{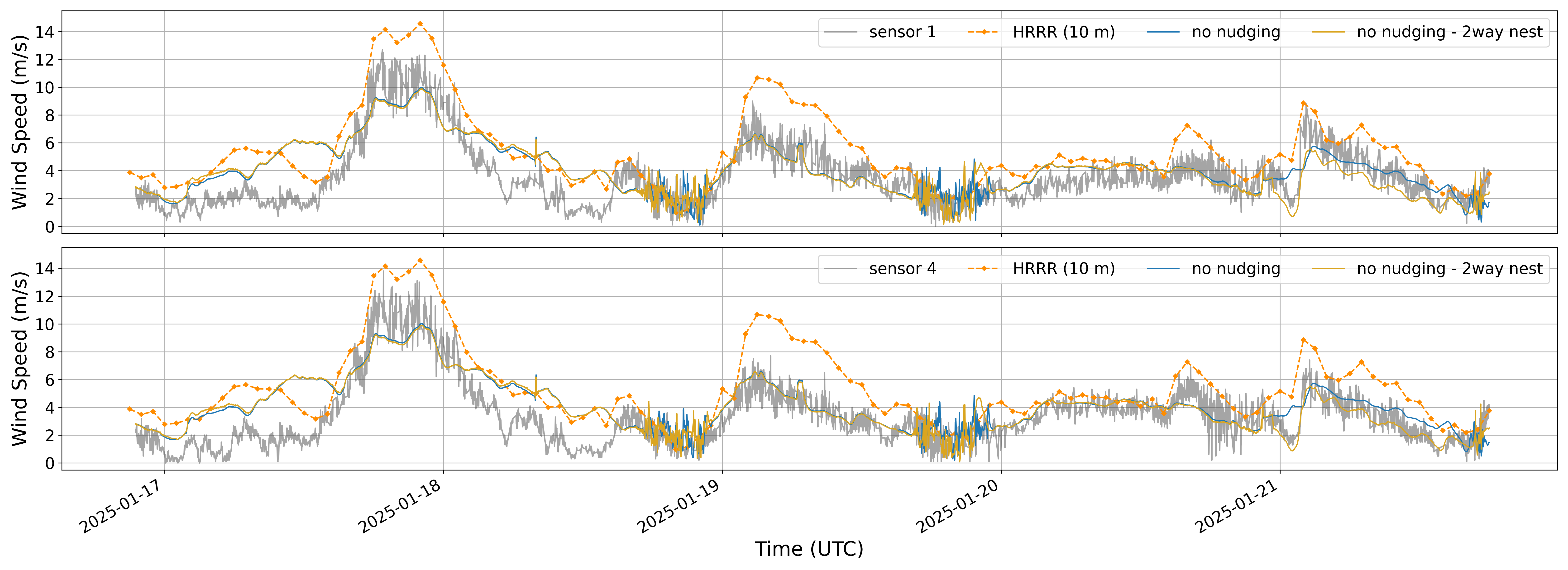}
\caption{Effect of two-way nesting without observational nudging. Near-surface wind-speed comparison from 16 January at 18:00 UTC through 21 January at 18:00 UTC. The upper and lower panels correspond to Sensors~1 and~4, respectively. Gray curves show the sensor measurements, and the colored curves show the d04 WRF solutions. The blue and gold curves corresponds to one- and two-way nesting sampled at the grid point nearest each sensor.}
\label{fig:timeSeries_Jan-2way}
\end{figure}

We first investigate the effect of two-way nesting for the case where the observational nudging is turned off. \Cref{fig:timeSeries_Jan-2way} compares the near-surface WRF wind speed with measurements at Sensors~1 and~4 during the January evaluation periods. The WRF has been setup with one-way and two-way nesting (blue and gold curves, respectively) while no observational nudging being enforced. We can see from the results that the two way nesting does not change the results much in the absence of observational nudging. We show later, however, that when we enforce the nudging, the two-way nesting setup provides a better agreement with the sensor measurement.

\subsection{Near-surface wind-speed response to observational nudging}

\begin{figure}[h]
\centering
\includegraphics[width=1\linewidth]{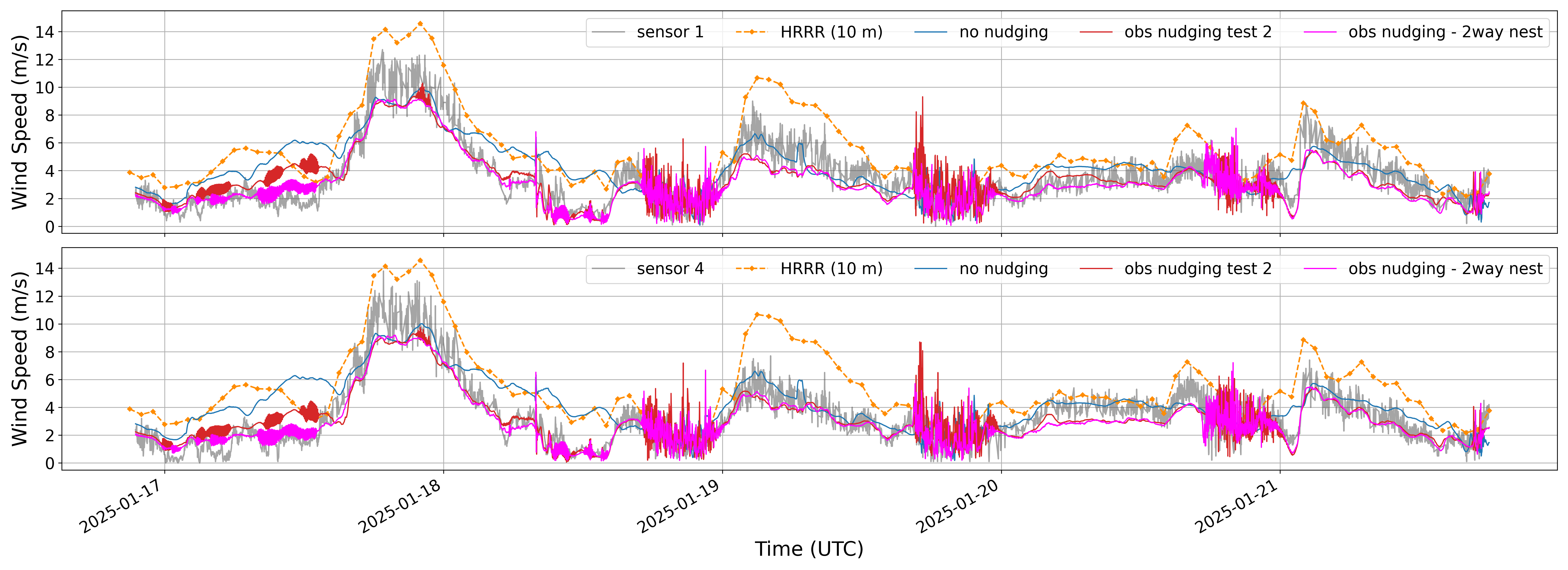}
\caption{Near-surface wind-speed comparison from 16 January at 18:00 UTC through 21 January at 18:00 UTC. The upper and lower panels correspond to Sensors~1 and~4, respectively. Gray curves show the sensor measurements, the orange curve shows hourly HRRR wind speed at $10~\mathrm{m}$, and the blue and red curves show the d04 WRF solutions without and with observational nudging. The magenta curve is for the case with observational nudging and two-way nesting. All the WRF simulation curves are sampled at the grid point nearest each sensor.}
\label{fig:timeSeries_Jan}
\end{figure}

\begin{figure}[h]
\centering
\includegraphics[width=1\linewidth]{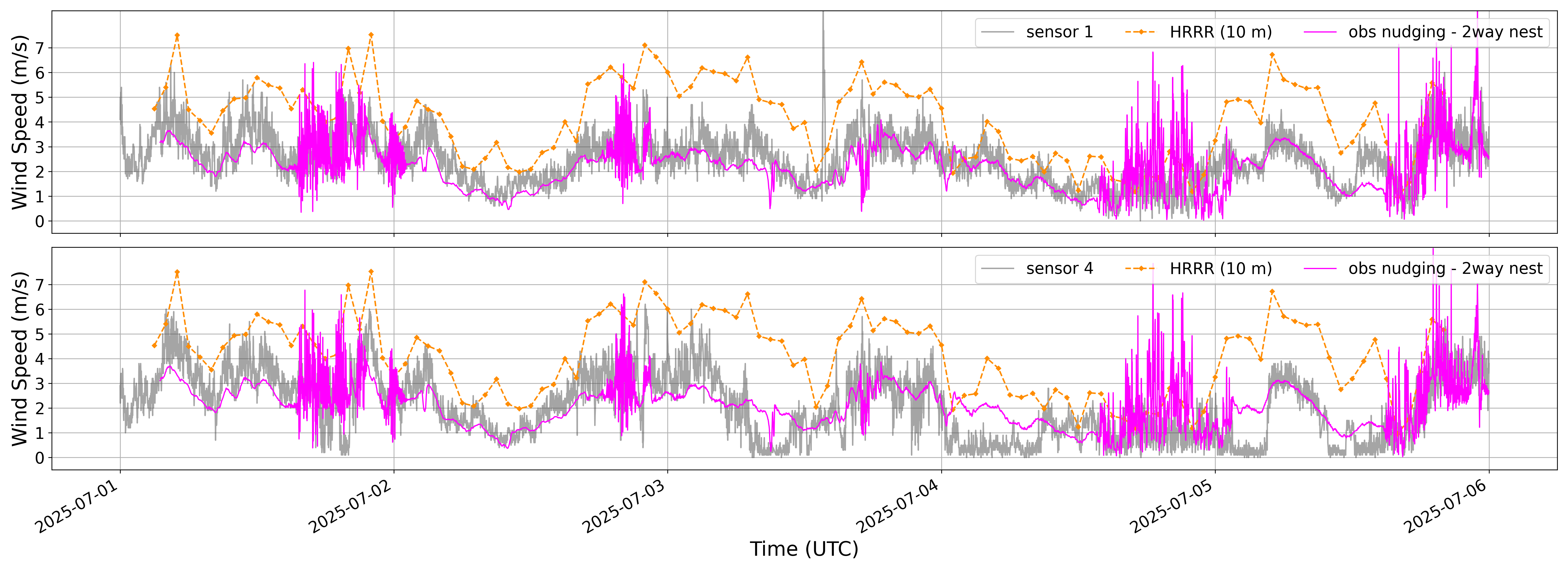}
\caption{Near-surface wind-speed comparison from 1 July at 00:00 UTC through 6 January at 00:00 UTC. The upper and lower panels correspond to Sensors~1 and~4, respectively. Gray curves show the sensor measurements, and the magenta curve shows the d04 WRF solutions with observational nudging and two-way nesting, sampled at the grid point nearest each sensor. }
\label{fig:timeSeries_Jul}
\end{figure}

\Cref{fig:timeSeries_Jan,fig:timeSeries_Jul} compare the near-surface WRF wind speed with measurements at Sensors~1 and~4 during the January and July evaluation periods. Both figures include the hourly HRRR wind at $10~\mathrm{m}$, providing a reference for the large-scale forcing from which the nested simulations are generated. HRRR captures the timing of the principal synoptic events but is generally smoother and faster than the near-surface observations. This difference is expected in part because HRRR represents a much larger horizontal area and a higher measurement level than the $1.5~\mathrm{m}$ anemometers, but it also illustrates why the coarse field cannot be used directly as the facility-scale wind.

The no-nudging d04 simulation improves the temporal resolution relative to HRRR and reproduces the broad sequence of wind-speed increases and decreases. In the January period, for example, it captures the onset and decay of the strong event spanning 17--18 January. However, the baseline run remains too smooth during several rapidly changing intervals and retains a positive bias during some weak-wind periods. Applying observational nudging shifts the d04 solution toward the local measurements at both sensors. The nudged run better follows the transition into and out of the major event and more rapidly approaches the observed low-wind states, although it still underestimates portions of the observed peak and does not reproduce every short-lived sensor fluctuation. The broadly similar response at Sensors~1 and~4 indicates that nudging corrects the site-scale background flow, whereas the remaining sensor-to-sensor differences likely include flow distortion from structures that are not represented in the WRF geometry.

The July period provides a complementary test under weaker and more intermittent winds. The no-nudging run frequently remains above the observed rolling mean, particularly at Sensor~4 during 3--5 July (not shown in the figure). Nudging reduces this persistent offset and improves the timing of several accelerations and calm periods. The instantaneous nudged series also contains substantially more short-time variability than the no-nudging series. This added variability makes the simulated signal more similar in character to the raw measurements, but it should not by itself be interpreted as independent evidence of correctly resolved turbulence because the same observations directly influence the nudged solution. The spectral and quantitative comparisons below are therefore used to distinguish improvements in the mean state from changes in variability and frequency content.


\subsection{Spectral content across the nested domains}

Spectral diagnostics complement pointwise error measures by examining how variability is distributed across scales. Skamarock \cite{Skamarock2004} used spatial kinetic-energy spectra to distinguish nominal grid spacing from effective model resolution and to assess the influence of numerical dissipation. Here, temporal wind-speed spectra provide a complementary diagnostic of frequency content at fixed locations; they are not a direct estimate of spatial effective resolution.

\begin{figure}[h]
\centering
\includegraphics[width=0.92\linewidth, trim=0cm 0cm 0cm 0cm, clip]{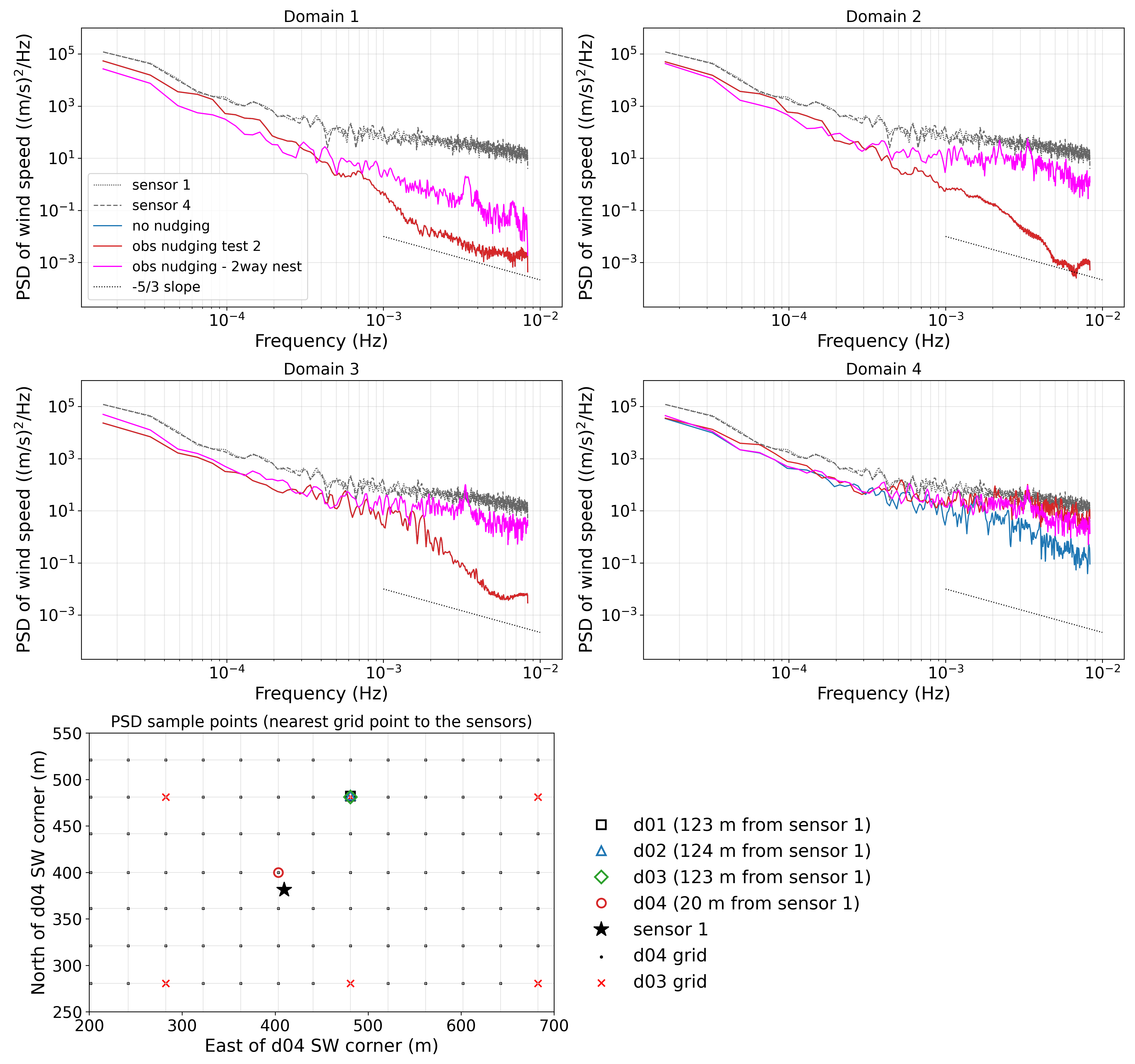}
\caption{Power spectral density of near-surface wind speed during the January 2025 evaluation period. Each panel shows one WRF domain together with the spectra from Sensors~1 and~4. The no-nudging solution is shown in blue; the observationally nudged solution is additionally shown in red for d04, where nudging is applied. The $-5/3$ line provides a reference spectral slope and is not normalized to the data.}
\label{fig:psd_Jan}
\end{figure}

The power spectral densities in \Cref{fig:psd_Jan} (and \Cref{fig:psd_Jul}) shows how the temporal variability changes as the HRRR-driven solution is refined from d01 to d04. At the lowest frequencies, the WRF and sensor spectra exhibit similar downward trends, indicating that the simulations retain the slowly varying synoptic and diurnal forcing. The principal difference among the domains appears at higher frequencies. The d01 and d02 spectra lose energy earliest, and d03 retains variability over a broader band before its spectrum also rolls off below the observations. This progressive displacement of the spectral roll-off toward higher frequency is consistent with the increasing spatial and temporal resolution of the nested domains.

The effect of observational nudging is concentrated in d04. Without nudging, the d04 spectrum contains more high-frequency energy than the parent domains but still falls increasingly below the sensor spectra toward the upper end of the resolved frequency range. The nudged d04 solution maintains substantially more power in this range and approaches the measured spectral level over a broader band in both the January and July cases. Thus, nudging changes more than the time-mean wind speed: it also modifies the distribution of temporal variance within the innermost domain. The $-5/3$ line is included only as a reference slope; its vertical placement is arbitrary, and agreement should be evaluated from the slope over a defined frequency interval rather than from the absolute position of the line.

The spectral comparison also identifies limits of the present result. The nudged spectrum does not coincide with both sensor spectra at every frequency, and the additional high-frequency power partly reflects the direct assimilation of one-minute observations. Consequently, the PSD results demonstrate that the nudged d04 field retains variability that is absent from the coarser and no-nudging simulations, but independent measurements or held-out observations are needed to determine how accurately that variability represents the spatial turbulence field away from the assimilated sensors.

In the one-way-nested configuration, observational nudging is applied only in d04, so the nudged and no-nudging solutions remain identical in the parent domains d01--d03. This behavior is evident in \Cref{fig:psd_Jan}, where the corresponding parent-domain spectra overlap, while the d04 spectra separate after the local observations are assimilated. With two-way nesting, however, the d04 correction is communicated back to the parent domains through the feedback mechanism described in \Cref{sec:2wnest}.

\begin{figure}[h]
\centering
\includegraphics[width=0.72\linewidth, trim=0cm 0cm 0cm 0cm, clip]{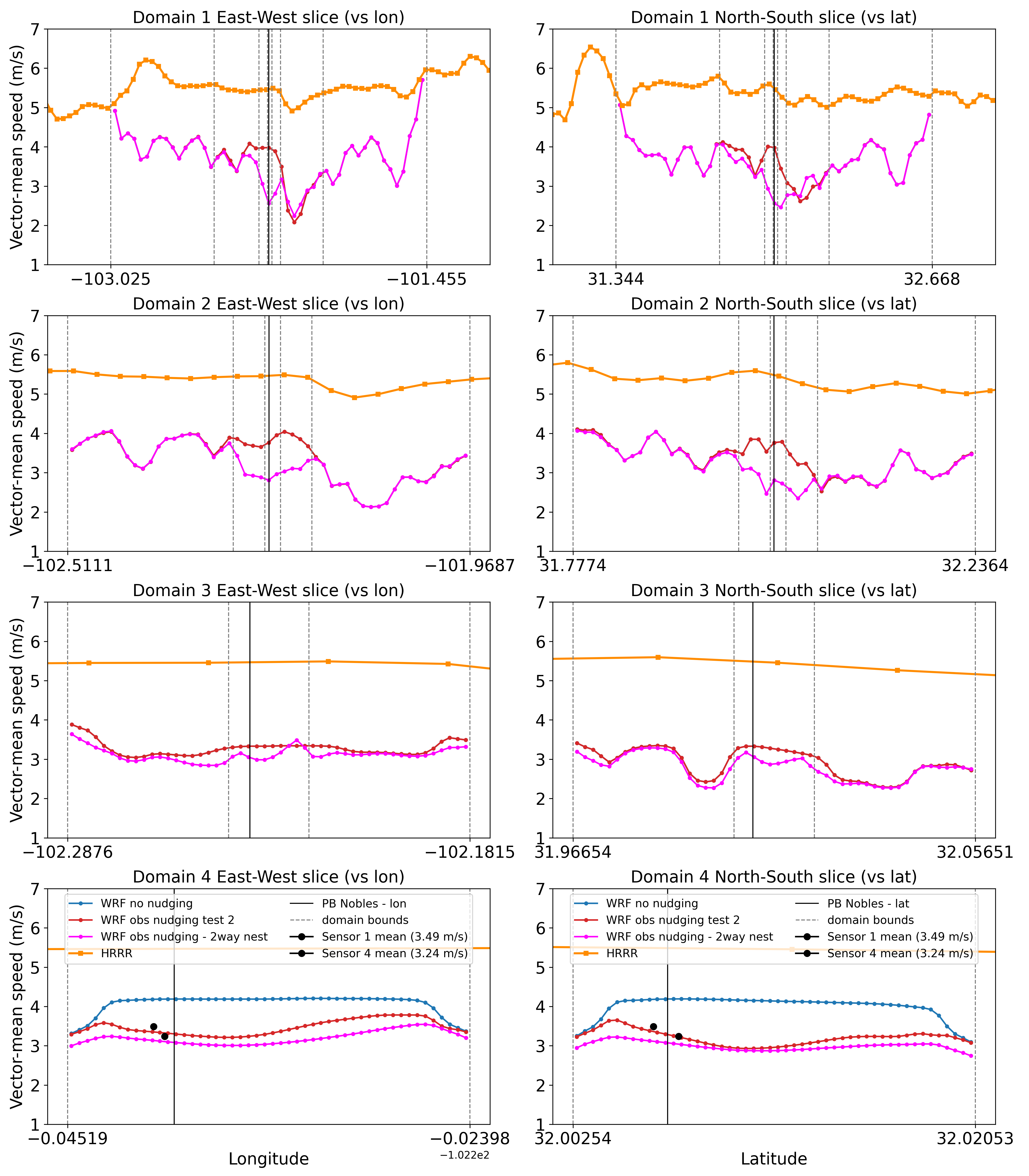}
\caption{East--west (left column) and north--south (right column) transects of the time-mean near-surface horizontal wind-vector magnitude during the 16--21 January 2025 evaluation period. Rows correspond to WRF domains d01--d04. Curves show the WRF solution without observational nudging (blue), the d04-nudged solution with one-way nesting (red), the observationally nudged solution with two-way nesting (magenta), and HRRR (orange). The solid black line marks the longitude or latitude of the PB-NOBLES-29H site, and dashed gray lines mark the nested-domain boundaries. In the one-way case, nudging changes d04 without altering its parent domains; in the two-way case, feedback propagates the d04 adjustment into d01--d03.}
\label{fig:slices_alld_Jan}
\end{figure}

The time-mean transects in \Cref{fig:slices_alld_Jan} demonstrate this distinction in physical space. Under one-way nesting, the red nudged solution overlaps the blue no-nudging solution in d01--d03 but differs substantially within d04, particularly along the north--south transect. The magenta two-way-nested solution separates from the one-way solutions in all four domains, showing that the site-scale observational adjustment has propagated upward through the nesting hierarchy. HRRR provides a comparatively smooth large-scale reference, whereas the WRF solutions develop progressively finer spatial variation toward d04. Thus, the choice of nesting feedback determines whether the influence of the assimilated site observations remains confined to the innermost domain or modifies the multiscale wind solution.

\subsection{Sensitivity to subfilter-scale closure and nesting feedback}

The \code{namelist} files in our test cases contain four observationally nudged WRF configurations. Tests~1--3 vary the d04 subfilter-scale option under one-way nesting, whereas Test~4 combines \texttt{sfs\_opt}=1 with two-way nesting. The tests use observations from Sensors~1 and~4 and otherwise share the d04 turbulence and observational-nudging parameters listed in \Cref{tab:wrf_sensitivity_tests}. This comparison is intended to determine whether the improvements attributed to nudging are robust to the turbulence closure and whether two-way communication between nested domains materially changes the site-scale solution. 

\begin{table}[htbp]
\centering
\caption{Configuration matrix transcribed from the four uploaded WRF namelists. Values for domain-dependent turbulence and observational-nudging options correspond to d04. Both on-site anemometers are used in all four tests.}
\label{tab:wrf_sensitivity_tests}
\scriptsize
\setlength{\tabcolsep}{4pt}
\renewcommand{\arraystretch}{1.25}
\begin{tabular}{lcccc}
\toprule
\textbf{Configuration parameter} & \textbf{Test 1} & \textbf{Test 2} & \textbf{Test 3} & \textbf{2way nest} \\
\midrule
Observations used & Sensors~1, 4 & Sensors~1, 4 & Sensors~1, 4 & Sensors~1, 4 \\
\texttt{feedback} & 0 (one-way) & 0 (one-way) & 0 (one-way) & 1 (two-way) \\
\texttt{sfs\_opt} (d04) & 0 & 1 & 2 & 1 \\
\bottomrule
\end{tabular}
\end{table}

\begin{figure}[h]
\centering
\includegraphics[width=1\linewidth]{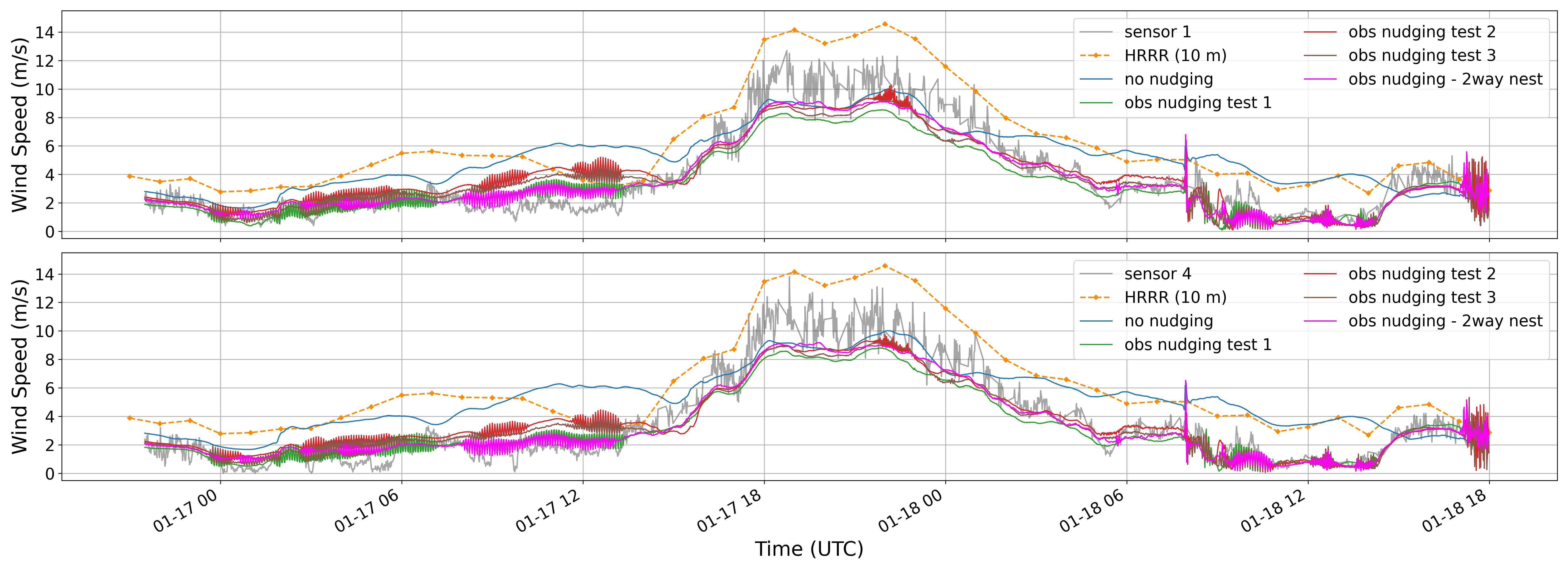}
\caption{Near-surface wind-speed sensitivity for 17--18 January 2025. The upper and lower panels correspond to Sensors~1 and~4. Gray curves show five-minute rolling-average observations, orange shows HRRR at $10~\mathrm{m}$, blue shows the no-nudging WRF baseline, and the remaining curves show observational-nudging tests~1--3 with alternative subfilter-scale and/or nesting configurations.}
\label{fig:timeSeries_Jan_test}
\end{figure}

The time series in \Cref{fig:timeSeries_Jan_test} show that all three nudged configurations respond more strongly to the observed changes than the no-nudging run. The largest common improvement occurs during low-wind periods, when the baseline remains elevated while the nudged simulations decrease toward the measurements. During the strong 17--18 January event, all configurations capture the rapid increase and subsequent decay, although each underestimates portions of the observed $10$--$12~\mathrm{m/s}$ peak. Test~1 generally produces the lowest wind speeds, while tests~2 and~3 remain close to one another and more closely reproduce the event amplitude over much of the peak and decay. The separation among the three nudged runs is nevertheless smaller than their collective separation from the no-nudging baseline, indicating that observation assimilation is the dominant control on the pointwise wind-speed response for this period.

\begin{figure}[h]
\centering
\includegraphics[width=0.75\linewidth, trim=0cm 0cm 0cm 1.5cm, clip]{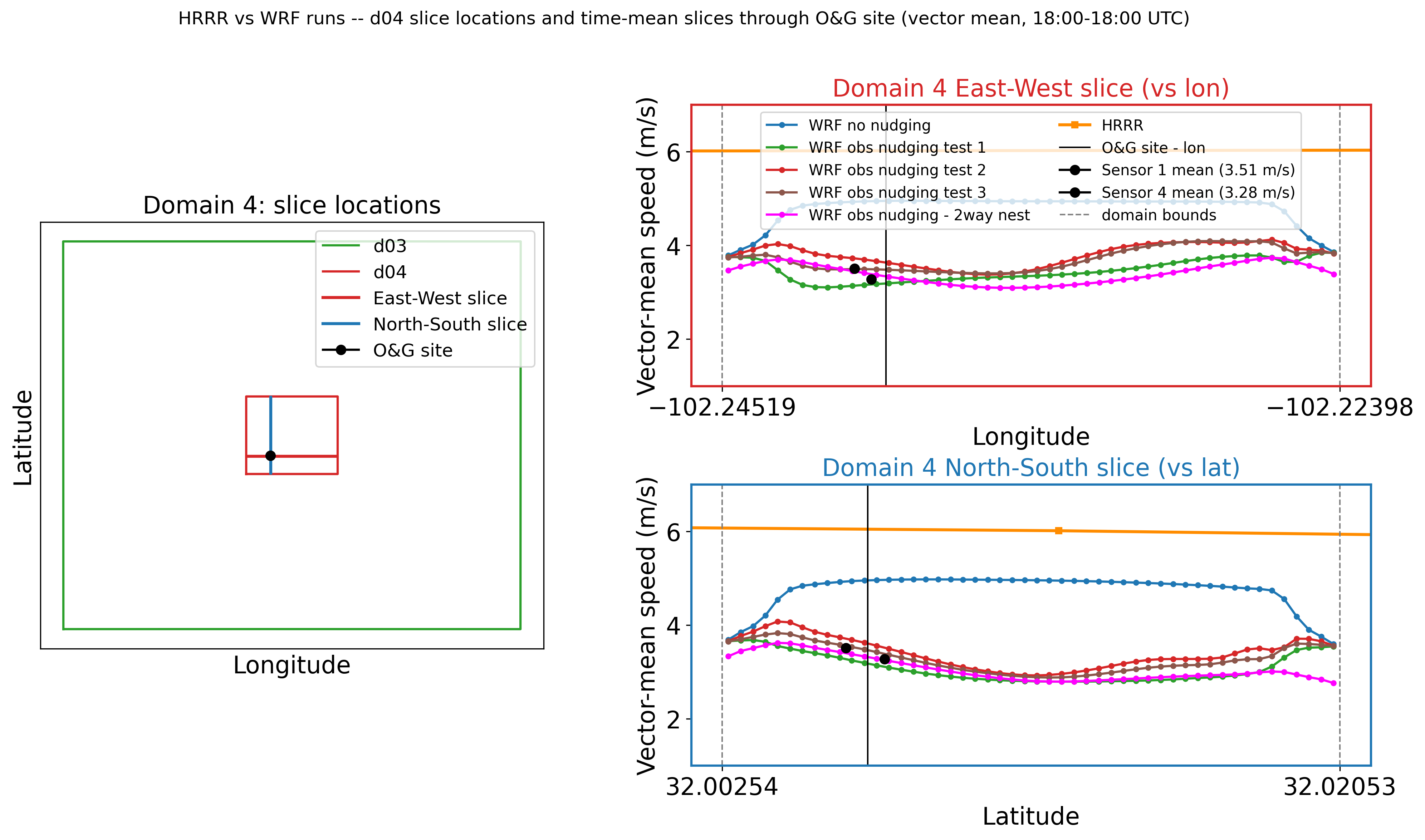}
\caption{Time-mean d04 wind-speed slices through PB-NOBLES-29H for the no-nudging baseline and observational-nudging tests~1--3. The left panel identifies the east--west and north--south transects. The right panels compare the corresponding profiles with HRRR; solid black lines mark the site coordinate and dashed gray lines mark the d04 boundaries.}
\label{fig:slices_Jan_test}
\end{figure}

The horizontal slices in \Cref{fig:slices_Jan_test} reveal differences that are less apparent in the point time series. The no-nudging solution is comparatively uniform across d04 and remains near $4~\mathrm{m/s}$ over most of both transects. The nudged cases reduce the mean speed near the site and introduce stronger spatial gradients, especially along the north--south transect. Test~1 produces the lowest and smoothest east--west profile, whereas tests~2 and~3 are similar near the facility but develop a deeper low-speed region farther along the north--south slice. These differences show that two configurations can agree at a sensor location while producing different wind fields elsewhere in the domain, which is important for plume-transport calculations that depend on the full spatial field rather than on a single point value.

\begin{figure}[h]
\centering
\includegraphics[width=1\linewidth, trim=0cm 12.5cm 0cm 0cm, clip]{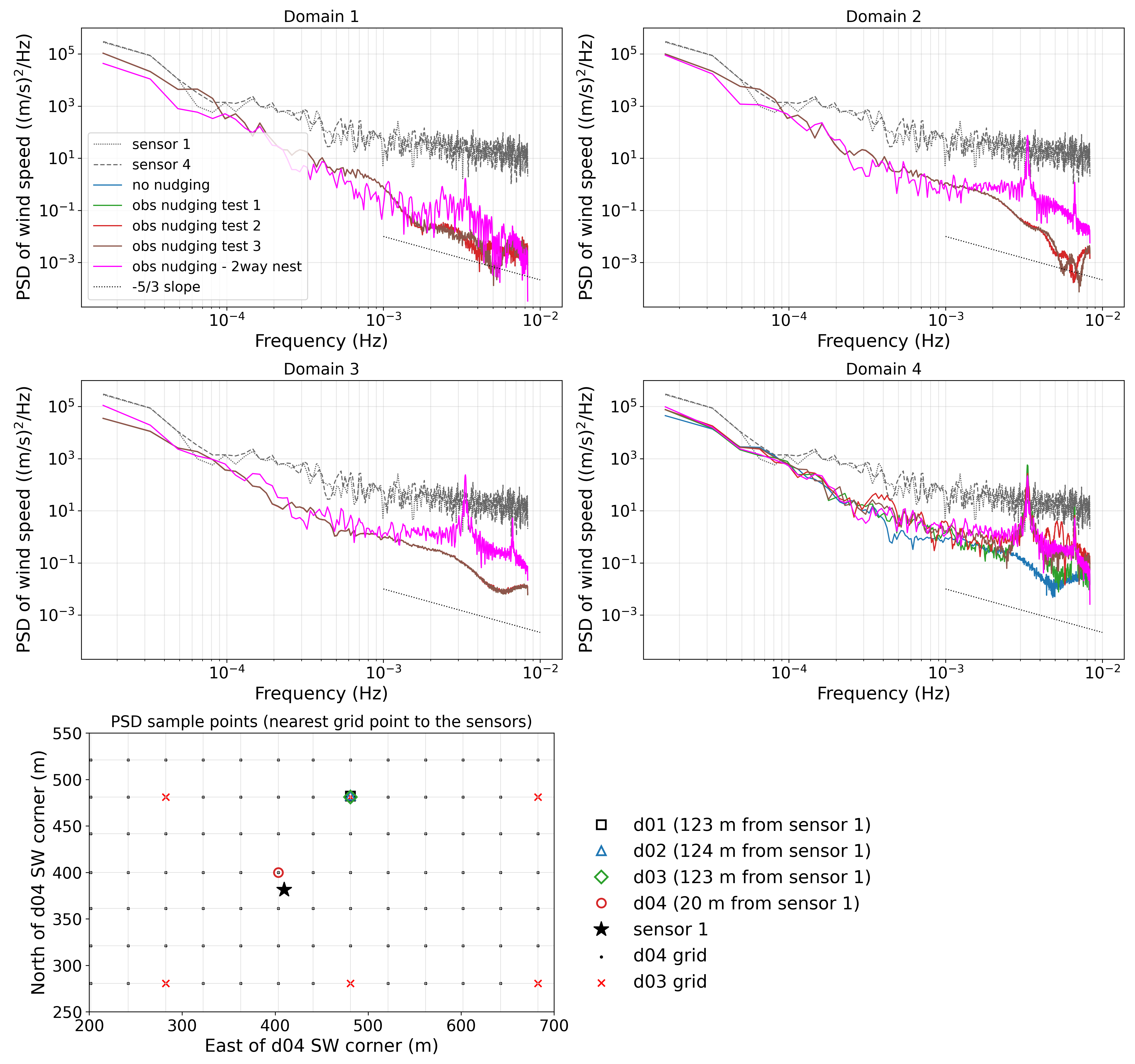}
\caption{Power spectral density for the January sensitivity experiment. The panels compare Sensors~1 and~4 with the no-nudging baseline and observational-nudging tests~1--3 in each WRF domain. The $-5/3$ line is shown as a reference slope. Differences among the sensitivity cases are concentrated primarily in d04.}
\label{fig:psd_Jan_test}
\end{figure}

The spectra in \Cref{fig:psd_Jan_test} provide a second view of this sensitivity. The curves are nearly indistinguishable in d01--d03 over the short analysis period, while the configurations separate clearly in d04 at intermediate and high frequencies. All three nudged cases retain more high-frequency energy than the no-nudging baseline, but the amount and distribution of this energy depend on the selected test. Tests~1 and~2 generally maintain more power over portions of the resolved high-frequency range, whereas test~3 is closer to the baseline at some frequencies. A narrow peak is also present in the nudged d04 spectra; before assigning it a physical interpretation, its frequency should be checked against the observation interval, model-output cadence, nudging window, and any processing filters. Taken together, the time series, spatial slices, and PSDs indicate that the model configuration has a secondary but non-negligible effect on the spatial gradients and small-scale variability generated after nudging.



\subsection{Quantitative comparison metrics}

\Cref{tab:metrics_s1_d4,tab:metrics_s4_d4} report the updated d04 metrics for 16--21 January 2025 at Sensors~1 and~4, respectively. Each table compares four configurations: no nudging with one-way nesting, no nudging with two-way nesting, d04 observational nudging with one-way nesting, and d04 observational nudging with two-way nesting. Unless otherwise specified, the baseline below is the no-nudging, one-way run. Accuracy, variability, and frequency metrics are sensor-specific; spatial, dynamics, and coherent-structure metrics describe the whole d04 field and are repeated in both tables. Definitions and interpretation are provided in \Cref{sec:appendix metric}.

\paragraph{Pointwise accuracy and nesting feedback.}
With one-way nesting, observational nudging reduces the RMSE from $1.594$ to $1.248~\mathrm{m\,s^{-1}}$ at Sensor~1 and from $1.717$ to $1.116~\mathrm{m\,s^{-1}}$ at Sensor~4. The corresponding MAE decreases from $1.223$ to $0.962~\mathrm{m\,s^{-1}}$ and from $1.305$ to $0.850~\mathrm{m\,s^{-1}}$. Bias changes from $+0.574$ to $-0.199~\mathrm{m\,s^{-1}}$ at Sensor~1 and from $+0.839$ to $-0.108~\mathrm{m\,s^{-1}}$ at Sensor~4: nudging removes much of the original positive bias but introduces a smaller negative bias. Correlation increases at both locations, from $0.740$ to $0.830$ and from $0.744$ to $0.875$, respectively.

Combining nudging with two-way nesting yields the lowest RMSE ($1.067$ and $0.957~\mathrm{m\,s^{-1}}$), lowest MAE ($0.828$ and $0.729~\mathrm{m\,s^{-1}}$), and highest correlation ($0.905$ and $0.925$) among the four configurations at Sensors~1 and~4. However, the biases become more negative ($-0.448$ and $-0.311~\mathrm{m\,s^{-1}}$), so one-way nudging retains the smallest absolute bias. Two-way nesting without nudging does not provide the same benefit: Sensor~1 RMSE increases from $1.594$ to $1.664~\mathrm{m\,s^{-1}}$, Sensor~4 RMSE remains $1.717~\mathrm{m\,s^{-1}}$, and correlation decreases at both sensors. The combined nudging--feedback configuration therefore improves overall pointwise agreement, but not every component of the error.

\paragraph{Variability and temporal persistence.}
All four configurations underestimate the observed speed variances of $4.846$ and $5.001~(\mathrm{m\,s^{-1}})^2$ at Sensors~1 and~4. One-way nudging reduces the variances from $3.134$ and $3.164$ to $2.967$ and $2.907~(\mathrm{m\,s^{-1}})^2$; two-way nudging raises them slightly to $3.070$ and $3.088~(\mathrm{m\,s^{-1}})^2$, still below the baseline and observations. In contrast, turbulence intensity increases from approximately $0.436$ at both sensors to $0.523$ and $0.545$ with one-way nudging, and to $0.576$ and $0.600$ with two-way nudging, approaching the observed values of $0.631$ and $0.691$. This distinction matters because intensity is normalized by mean speed: it can improve even while absolute variance remains too small. Integral time scales also move closer to the observations. The baseline values of approximately $14{,}000~\mathrm{s}$ decrease to approximately $10{,}400$ and $9{,}750~\mathrm{s}$ with one-way nudging, compared with observed values of $9429.9$ and $9912.1~\mathrm{s}$. The two-way-nudged values are $10444.7$ and $9984.0~\mathrm{s}$. These metrics indicate improved temporal persistence, particularly at Sensor~4, but not recovery of the full observed fluctuation amplitude.

\paragraph{Frequency content.}
Spectral error decreases at both sensors: from $0.598$ and $0.584$ in the baseline to $0.511$ and $0.446$ with one-way nudging, and to $0.478$ and $0.410$ with two-way nudging. Band-averaged coherence also increases, from $0.038$ and $0.029$ to $0.051$ and $0.059$, and then to $0.054$ and $0.061$, respectively. Although these changes are favorable, coherence remains low and should not be equated with the much higher time-domain correlation. Spectral slopes do not improve uniformly: both sensors have a fitted slope of $-1.67$, whereas the one-way- and two-way-nudged slopes are $-1.88$ and $-1.97$ at Sensor~1, and $-2.03$ and $-2.04$ at Sensor~4. Thus, reduced integrated spectral error does not imply that the spectral shape or slope is reproduced at every frequency.

\paragraph{Spatial structure and flow gradients.}
Relative to the baseline, one-way nudging increases domain-wide spatial variance from $0.332$ to $1.154~(\mathrm{m\,s^{-1}})^2$; two-way nudging gives an intermediate value of $0.949~(\mathrm{m\,s^{-1}})^2$. RMS vorticity similarly rises from $3.08\times10^{-3}$ to $7.59\times10^{-3}~\mathrm{s^{-1}}$ with one-way nudging and to $4.28\times10^{-3}~\mathrm{s^{-1}}$ with two-way nudging. Enstrophy and the reported RMS shear diagnostic follow the same ordering. The leading POD-mode fraction decreases from $0.979$ to $0.811$ with one-way nudging and to $0.856$ with two-way nudging, indicating a redistribution of field variance across more spatial modes. These changes describe increased spatial heterogeneity and stronger resolved gradients, not independently demonstrated improvements in spatial accuracy. The $1000~\mathrm{m}$ correlation lengths and approximately $2.5$--$2.7~\mathrm{km}$ spatial spectral peaks approach or exceed the scales accessible within d04 and should be treated as potentially domain-limited diagnostics. The very low dominant DMD frequency in the two-way-nudged run ($2.37\times10^{-6}~\mathrm{Hz}$) corresponds to a period of about $4.9$ days, nearly the full evaluation interval; it does not establish a well-sampled periodic structure.

Overall, the updated January metrics support observational nudging as the main source of improved pointwise agreement, with two-way feedback providing additional reductions in RMSE, MAE, and spectral error. This benefit coexists with greater negative bias than in one-way nudging, underestimated temporal variance, and imperfect spectral slopes. Because both sensors directly constrain the nudged solutions, independent or held-out measurements remain necessary to assess accuracy away from the assimilated locations.

\begin{table}[htbp]
\centering
\caption{Wind metrics for 16--21 January 2025 at Sensor~1 and across WRF domain d04. ``No Nudge'' and ``Nudge d04'' denote one-way nesting unless ``2-Way Nest'' is specified. Accuracy, variability, and frequency metrics are evaluated at the sensor location; spatial, dynamics, and coherent-structure metrics are domain-wide and are repeated in \Cref{tab:metrics_s4_d4}. Both sensors are assimilated in the nudged runs, so these are not independent validation statistics. Definitions and caveats are given in \Cref{sec:appendix metric}; ``--'' denotes a quantity not applicable or unavailable for the observations.}
\label{tab:metrics_s1_d4}
\resizebox{\textwidth}{!}{%
\begin{tabular}{llrrrrr}
\toprule
Category & Metric & Sensor 1 & No Nudge & No Nudge 2-Way Nest & Nudge d04 & Nudge d04 2-Way Nest \\
\midrule
Accuracy & RMSE (m/s) & -- & 1.594 & 1.664 & 1.248 & 1.067 \\
 & MAE (m/s) & -- & 1.223 & 1.310 & 0.962 & 0.828 \\
 & Bias (m/s) & -- & 0.574 & 0.435 & -0.199 & -0.448 \\
 & Correlation & -- & 0.740 & 0.699 & 0.830 & 0.905 \\
\midrule
Variability & Variance ((m/s)$^2$) & 4.846 & 3.134 & 3.441 & 2.967 & 3.070 \\
 & Turbulence Intensity & 0.631 & 0.436 & 0.473 & 0.523 & 0.576 \\
 & Integral Time Scale (s) & 9429.9 & 13992.9 & 14924.3 & 10432.5 & 10444.7 \\
\midrule
Frequency & Spectral Slope & -1.67 & -1.96 & -1.98 & -1.88 & -1.97 \\
 & Spectral Error & -- & 0.598 & 0.546 & 0.511 & 0.478 \\
 & Coherence & -- & 0.038 & 0.035 & 0.051 & 0.054 \\
\midrule
Spatial & Correlation Length (m) & -- & 1000 & 1000 & 1000 & 1000 \\
 & Spatial Variance ((m/s)$^2$) & -- & 0.332 & 0.297 & 1.154 & 0.949 \\
 & Spatial Spectral Peak (m) & -- & 2577 & 2524 & 2583 & 2690 \\
\midrule
Dynamics & RMS Vorticity (s$^{-1}$) & -- & 3.08e-03 & 2.88e-03 & 7.59e-03 & 4.28e-03 \\
 & Max Vorticity (s$^{-1}$) & -- & 9.12e-02 & 1.46e-01 & 1.89e-01 & 1.08e-01 \\
 & Enstrophy (s$^{-2}$) & -- & 5.48e-06 & 5.23e-06 & 5.20e-05 & 1.49e-05 \\
 & Mean Divergence (s$^{-1}$) & -- & 1.31e-04 & 2.82e-04 & 1.29e-04 & -9.27e-05 \\
 & RMS Shear (s$^{-1}$) & -- & 3.26e-03 & 2.91e-03 & 6.88e-03 & 4.52e-03 \\
\midrule
Coherent Structures & POD Mode-1 Energy & -- & 0.979 & 0.975 & 0.811 & 0.856 \\
 & POD Modes 1--5 Energy & -- & 0.990 & 0.991 & 0.936 & 0.951 \\
 & Dominant DMD Frequency (Hz) & -- & 3.09e-05 & 1.14e-05 & 1.01e-05 & 2.37e-06 \\
\bottomrule
\end{tabular}
}
\end{table}

\begin{table}[htbp]
\centering
\caption{Wind metrics for 16--21 January 2025 at Sensor~4 and across WRF domain d04. ``No Nudge'' and ``Nudge d04'' denote one-way nesting unless ``2-Way Nest'' is specified. Accuracy, variability, and frequency metrics are evaluated at the sensor location; domain-wide spatial, dynamics, and coherent-structure metrics are identical to those in \Cref{tab:metrics_s1_d4}. Both sensors are assimilated in the nudged runs, so these are not independent validation statistics. Definitions and caveats are given in \Cref{sec:appendix metric}; ``--'' denotes a quantity not applicable or unavailable for the observations.}
\label{tab:metrics_s4_d4}
\resizebox{\textwidth}{!}{%
\begin{tabular}{llrrrrr}
\toprule
Category & Metric & Sensor 4 & No Nudge & No Nudge 2-Way Nest & Nudge d04 & Nudge d04 2-Way Nest \\
\midrule
Accuracy & RMSE (m/s) & -- & 1.717 & 1.717 & 1.116 & 0.957 \\
 & MAE (m/s) & -- & 1.305 & 1.272 & 0.850 & 0.729 \\
 & Bias (m/s) & -- & 0.839 & 0.697 & -0.108 & -0.311 \\
 & Correlation & -- & 0.744 & 0.721 & 0.875 & 0.925 \\
\midrule
Variability & Variance ((m/s)$^2$) & 5.001 & 3.164 & 3.437 & 2.907 & 3.088 \\
 & Turbulence Intensity & 0.691 & 0.436 & 0.471 & 0.545 & 0.600 \\
 & Integral Time Scale (s) & 9912.1 & 14153.6 & 15135.5 & 9749.7 & 9984.0 \\
\midrule
Frequency & Spectral Slope & -1.67 & -2.00 & -2.07 & -2.03 & -2.04 \\
 & Spectral Error & -- & 0.584 & 0.534 & 0.446 & 0.410 \\
 & Coherence & -- & 0.029 & 0.038 & 0.059 & 0.061 \\
\midrule
Spatial & Correlation Length (m) & -- & 1000 & 1000 & 1000 & 1000 \\
 & Spatial Variance ((m/s)$^2$) & -- & 0.332 & 0.297 & 1.154 & 0.949 \\
 & Spatial Spectral Peak (m) & -- & 2577 & 2524 & 2583 & 2690 \\
\midrule
Dynamics & RMS Vorticity (s$^{-1}$) & -- & 3.08e-03 & 2.88e-03 & 7.59e-03 & 4.28e-03 \\
 & Max Vorticity (s$^{-1}$) & -- & 9.12e-02 & 1.46e-01 & 1.89e-01 & 1.08e-01 \\
 & Enstrophy (s$^{-2}$) & -- & 5.48e-06 & 5.23e-06 & 5.20e-05 & 1.49e-05 \\
 & Mean Divergence (s$^{-1}$) & -- & 1.31e-04 & 2.82e-04 & 1.29e-04 & -9.27e-05 \\
 & RMS Shear (s$^{-1}$) & -- & 3.26e-03 & 2.91e-03 & 6.88e-03 & 4.52e-03 \\
\midrule
Coherent Structures & POD Mode-1 Energy & -- & 0.979 & 0.975 & 0.811 & 0.856 \\
 & POD Modes 1--5 Energy & -- & 0.990 & 0.991 & 0.936 & 0.951 \\
 & Dominant DMD Frequency (Hz) & -- & 3.09e-05 & 1.14e-05 & 1.01e-05 & 2.37e-06 \\
\bottomrule
\end{tabular}
}
\end{table}

\section{Summary and Conclusions} \label{sec:Conclusions}

This study addresses a central limitation in continuous methane monitoring at oil and gas facilities: plume transport and source inversion require a spatially and temporally varying wind field, whereas operational weather products are too coarse for direct facility-scale use and on-site measurements are available at only a few points. We developed a physics-based multiscale WRF framework for the PB-NOBLES-29H facility in the Permian Basin. Hourly HRRR fields at $3~\mathrm{km}$ resolution were dynamically downscaled through four nested domains with grid spacings of $3~\mathrm{km}$, $1~\mathrm{km}$, $200~\mathrm{m}$, and $40~\mathrm{m}$. The outer domains represented mesoscale boundary-layer evolution using PBL parameterization, while the inner domains operated in LES mode. One-minute measurements from two near-surface anemometers were assimilated through wind-only observational nudging in d04, and the effects of nesting feedback and alternative subfilter-scale configurations were examined.

The baseline WRF simulation retained the timing of the principal synoptic events and generated substantially finer temporal and spatial structure than HRRR. It reproduced the broad increases and decreases in wind speed during both the January and July 2025 periods, but it remained too smooth during rapidly changing conditions and exhibited a positive bias during several weak-wind intervals. Observational nudging produced the clearest and most consistent improvement in the near-surface mean wind. For the January evaluation period with one-way nesting, the wind-speed bias decreased from $0.574$ to $-0.199~\mathrm{m\,s^{-1}}$ at Sensor~1 and from $0.839$ to $-0.108~\mathrm{m\,s^{-1}}$ at Sensor~4. The corresponding RMSE decreased from $1.594$ to $1.248~\mathrm{m\,s^{-1}}$ and from $1.717$ to $1.116~\mathrm{m\,s^{-1}}$, while the MAE also decreased at both locations. The nudged simulations more readily approached observed calm conditions and generally followed the transitions into and out of major wind events more closely than the no-nudging simulation. These results establish observational nudging as a critical component of the present framework: high spatial resolution alone did not remove the site-scale bias inherited from the model forcing and configuration.

Nudging also changed the character of the resolved flow. The d04 spectra retained more high-frequency energy, and the spatial wind fields exhibited stronger gradients and less uniformity than the no-nudging solution. Domain-wide spatial variance increased from $0.332$ to $1.154~(\mathrm{m\,s^{-1}})^2$, accompanied by increases in RMS vorticity, enstrophy, and shear. The sensitivity experiments further showed that alternative nesting and subfilter-scale configurations can modify spatial gradients and high-frequency variability even when their wind speeds are similar at an observation point. Nevertheless, the separation among the nudged sensitivity cases was generally smaller than their collective separation from the no-nudging baseline, indicating that assimilation of the observations was the dominant control on pointwise wind-speed improvement. Two-way nesting alone produced little change when nudging was disabled; its greater relevance is in allowing the dynamically adjusted fine-domain state to influence the parent domains when observational nudging is active.

The improvements are not uniform across every metric. Temporal correlation increased at both sensors, while variance remained underestimated and changes in spectral slope depended on location. The additional high-frequency energy and spatial structure therefore should not automatically be interpreted as more accurate turbulence. Both anemometers were used in the nudging procedure, so agreement at those locations measures the model's assimilation response rather than providing independent validation of the reconstructed field. In addition, the WRF geometry does not explicitly resolve tanks and other facility equipment that can shelter or redirect the flow near the sensors. Independent or held-out observations---preferably including measurements at additional locations, heights, or along spatial transects---are needed to determine the accuracy of the wind field away from the assimilated instruments.

More broadly, the work confirms that producing credible wind fields at tens-of-meters resolution is not a routine matter of increasing the WRF grid resolution. The simulation must bridge mesoscale, gray-zone, and LES regimes; select compatible turbulence, surface, and land-surface treatments; provide adequate vertical resolution; manage nesting and feedback; and allow the fine domains sufficient spin-up and spatial fetch. These challenges are consistent with the multiscale modeling literature, which emphasizes the sensitivity of high-resolution simulations to forcing, land-surface representation, turbulence treatment, grid design, and turbulence initialization \cite{Talbot2012,Haupt2019MMC,Haupt2023,GianiCrippa2024,AlOqaily2025}. The computational burden is also substantial: in the present configuration, a 24-hour simulation required approximately 12--18 hours of wall-clock time on 12--20 high-performance-computing nodes. The effort required to configure, execute, diagnose, and validate these simulations is therefore incompatible with the rapid turnaround needed for routine methane-monitoring operations.

The high-resolution WRF fields developed here are intended to bridge that operational gap by serving as physics-based training and evaluation data for fast surrogate models. A surrogate can learn the mapping from coarse HRRR forcing and sparse local anemometer measurements to a continuous site-scale wind field without repeating the full multiscale WRF calculation for every monitoring period. Such a model could provide the rapid wind-field estimates required for methane plume simulation, source localization, and emission-rate estimation, while retaining information learned from physically consistent mesoscale-to-LES simulations. The present work thus provides both a demonstrated pathway for improving site-scale winds through observational nudging and the data foundation for a computationally efficient methane-monitoring workflow. Continued progress will require independent spatial validation and careful assessment of how accurately the surrogate preserves the mean flow, temporal variability, and spatial structure that are important for plume transport.

\section*{Acknowledgements}
\noindent This research was supported by ExxonMobil Technology and Engineering Company. 
The authors gratefully acknowledge Dr.~Aniruddha Bora of Texas State University and Dr.~Sheng-Lun Tai of Pacific Northwest National Laboratory for their support, insightful discussions, and constructive suggestions throughout the development of this work.
Any opinions, findings, conclusions, or recommendations are those of the authors and not of the sponsors.

\appendix
\newpage

\section{Additional WRF results} \label{sec:appendix add results}


\begin{figure}[h]
\centering
\includegraphics[width=0.8\linewidth,trim=0cm 12.5cm 0cm 0.3cm, clip]{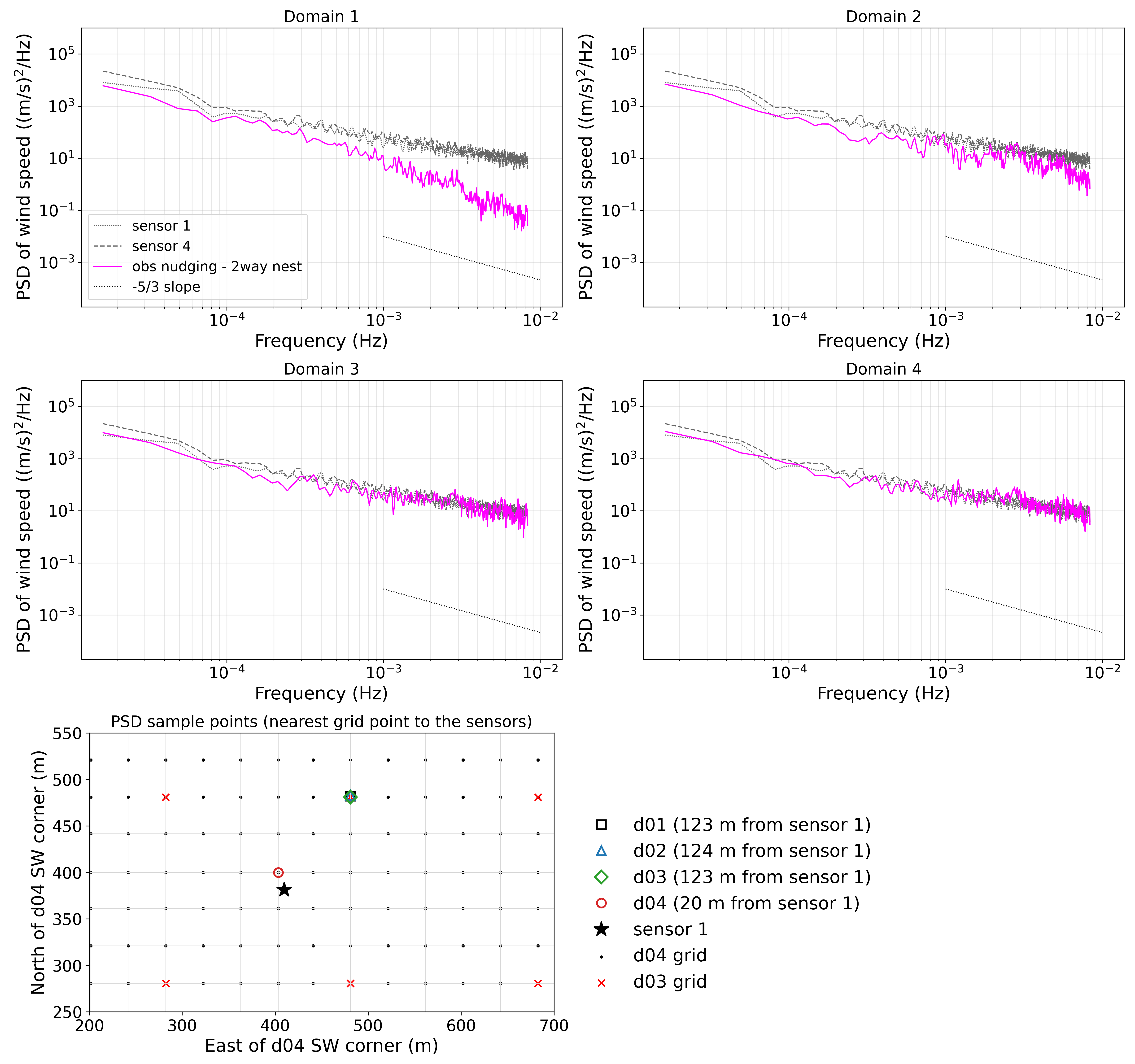}
\caption{Power Spectral Density plots of the simulated (WRF) vs sensor data for each simulation domain. Simulation period is July 1 - 6}
\label{fig:psd_Jul}
\end{figure}


\section{Namelist templates for controlled sensitivity tests}

\subsection{Baseline from Table~1: 1.5-order TKE}

\begin{lstlisting}
&physics
 bl_pbl_physics = 1, 0, 0, 0,
/

&dynamics
 diff_opt = 2, 2, 2, 2,
 km_opt   = 4, 2, 2, 2,
 sfs_opt  = 0, 0, 0, 0,
/
\end{lstlisting}

\subsection{Diagnostic NBA1 with Smagorinsky scalar mixing}

\begin{lstlisting}
&physics
 bl_pbl_physics = 1, 0, 0, 0,
/

&dynamics
 diff_opt = 2, 2, 2, 2,
 km_opt   = 4, 3, 3, 3,
 sfs_opt  = 0, 1, 1, 1,
/
\end{lstlisting}

Here NBA1 supplies momentum stresses on d02--d04 and
\code{km\_opt=3} supplies scalar diffusivities.

\subsection{Diagnostic NBA1 with TKE-based scalar mixing}

\begin{lstlisting}
&physics
 bl_pbl_physics = 1, 0, 0, 0,
/

&dynamics
 diff_opt = 2, 2, 2, 2,
 km_opt   = 4, 2, 2, 2,
 sfs_opt  = 0, 1, 1, 1,
/
\end{lstlisting}

Here NBA1 supplies momentum stresses while the prognostic-TKE pathway
supplies scalar diffusivities.

\subsection{TKE-based NBA2}

\begin{lstlisting}
&physics
 bl_pbl_physics = 1, 0, 0, 0,
/

&dynamics
 diff_opt = 2, 2, 2, 2,
 km_opt   = 4, 2, 2, 2,
 sfs_opt  = 0, 2, 2, 2,
/
\end{lstlisting}

Here the prognostic SGS TKE is part of both the NBA2 momentum stress
and the scalar mixing closure.

\section{Metric definitions and interpretation} \label{sec:appendix metric}

The metrics in \Cref{tab:metrics_s1_d4,tab:metrics_s4_d4} describe complementary aspects of model performance. Accuracy metrics compare simulated and observed wind speeds directly; variability and spectral metrics compare their statistical behavior; spatial, dynamics, and coherent-structure metrics describe the simulated d04 field. The latter have no corresponding spatial observations in this study, so larger values do not, by themselves, indicate greater accuracy. Because both anemometers are assimilated, agreement at their locations measures the assimilation response rather than independent validation.

Let $o_i$ and $s_i$ denote time-aligned observed and simulated wind speeds, with $N$ valid paired samples. For a generic scalar wind-speed series $q_i$, $\bar q$ and $\sigma_q$ denote its mean and population standard deviation. Horizontal velocity components are denoted by $u$ and $v$; these should not be confused with scalar wind speed. Spatial averages are written as $\langle\cdot\rangle_{x,y}$ and averages over $N_f$ model snapshots as $\langle\cdot\rangle_{\mathrm{frames}}$. Comparisons require consistent sampling, averaging, detrending, and treatment of missing data. In particular, variability over a multiday record includes meteorological changes as well as turbulent fluctuations.

\subsection{Accuracy metrics}

\paragraph{Root-mean-square error (RMSE)}
\begin{equation}
\mathrm{RMSE}=\sqrt{\frac{1}{N}\sum_{i=1}^{N}(s_i-o_i)^2}.
\end{equation}
RMSE measures the typical error magnitude in $\mathrm{m\,s^{-1}}$, giving relatively greater weight to large errors. Smaller values indicate better agreement, with zero denoting an exact match. It combines systematic bias with errors in the timing and amplitude of fluctuations.

\paragraph{Mean absolute error (MAE)}
\begin{equation}
\mathrm{MAE}=\frac{1}{N}\sum_{i=1}^{N}|s_i-o_i|.
\end{equation}
MAE is the average unsigned speed error in $\mathrm{m\,s^{-1}}$. It is less sensitive to occasional large errors than RMSE; a large separation between RMSE and MAE suggests that relatively large departures contribute appreciably to the error. Both metrics should decrease for improved pointwise agreement.

\paragraph{Bias}
\begin{equation}
\mathrm{Bias}=\frac{1}{N}\sum_{i=1}^{N}(s_i-o_i)=\bar s-\bar o.
\end{equation}
Positive bias indicates overprediction of mean speed and negative bias indicates underprediction. Improvement means movement toward zero, not simply a decrease in the signed value. A small bias can coexist with substantial RMSE because positive and negative errors cancel in the average.

\paragraph{Correlation}
\begin{equation}
r=\frac{\sum_{i=1}^{N}(o_i-\bar o)(s_i-\bar s)}
{\sqrt{\sum_{i=1}^{N}(o_i-\bar o)^2}\sqrt{\sum_{i=1}^{N}(s_i-\bar s)^2}}.
\end{equation}
Pearson correlation ranges from $-1$ to $1$ and describes linear covariation. Values near $1$ indicate that simulated and observed speeds rise and fall together, but do not ensure correct mean speed or fluctuation amplitude. Correlation is undefined for a constant series and should be interpreted alongside bias, RMSE, and variance.

\paragraph{Standard-deviation ratio}
\begin{equation}
R_\sigma=\frac{\sigma_s}{\sigma_o}.
\end{equation}
A ratio of one indicates matching variability amplitude; values below or above one indicate under- or overdispersion, respectively. Matching standard deviations does not imply matching time evolution. This ratio is included for interpretation but is not a separate row in the updated tables.

\subsection{Variability metrics}

\paragraph{Variance}
\begin{equation}
\sigma_q^2=\frac{1}{N}\sum_{i=1}^{N}(q_i-\bar q)^2.
\end{equation}
Variance, in $(\mathrm{m\,s^{-1}})^2$, measures the amplitude of wind-speed fluctuations about the record mean. Agreement with the observed variance is the relevant target, rather than maximizing the simulated value. This scalar speed variance is not turbulent kinetic energy, which requires fluctuations of the velocity components and a specified mean-flow separation.

\paragraph{Turbulence intensity (TI)}
\begin{equation}
\mathrm{TI}=\frac{\sigma_q}{\bar q}.
\end{equation}
TI is dimensionless and measures fluctuation amplitude relative to mean speed; a value of $0.5$ means that the standard deviation is half the mean. It can increase when the mean decreases even if the variance also decreases. Consequently, improved TI alone does not establish improved absolute variability. For these multiday speed records, the reported TI includes low-frequency meteorological variability and is not a turbulence-only statistic; it also becomes unstable when the mean approaches zero.

\paragraph{Gust factor}
\begin{equation}
G=\frac{\max_i q_i}{\bar q}.
\end{equation}
The gust factor expresses the largest sampled speed relative to the mean. It is sensitive to sampling cadence, averaging interval, record length, and isolated extremes, so comparisons require consistent processing. It is included here as a supplementary definition but is not reported in the updated tables.

\paragraph{Integral time scale}
For $q'_i=q_i-\bar q$, define the normalized autocorrelation at integer sample lag $\ell$ by
\begin{equation}
\rho_\ell=\frac{\sum_{i=1}^{N-\ell}q'_i q'_{i+\ell}}{\sum_{i=1}^{N}(q'_i)^2},
\qquad t_\ell=\ell\Delta t,\qquad \Delta t=60~\mathrm{s}.
\end{equation}
The integral time scale is the area under the positive autocorrelation lobe up to its first zero crossing $t_0$:
\begin{equation}
\mathcal{T}=\int_0^{t_0}\rho(t)\,dt.
\end{equation}
Numerical integration in sample-lag coordinates requires multiplication by $\Delta t$; integration in physical seconds does not require that factor again. Larger $\mathcal{T}$ indicates more persistent fluctuations, whereas smaller values indicate faster decorrelation. The aim is agreement with the observed time scale, not its minimization. Over a multiday record, sustained weather events can dominate this measure; it should not automatically be interpreted as an eddy lifetime. If the autocorrelation never crosses zero within the available lags, the estimate is record-limited rather than a well-resolved decorrelation time.

\subsection{Spectral metrics}

The power spectral density (PSD), $S(f)$, describes variance per unit frequency and has units of $(\mathrm{m\,s^{-1}})^2\,\mathrm{Hz}^{-1}$. The existing spectral-estimation convention uses Welch averaging with linear detrending, segment length $\min(1024,N)$, and sampling frequency $f_s=1/60~\mathrm{Hz}$. Segment length, window, overlap, and the fitted frequency band must be consistent when comparing spectra. At one-minute sampling, frequencies above the Nyquist frequency $1/120~\mathrm{Hz}$ cannot be resolved.

\paragraph{Spectral slope}
\begin{equation}
\log_{10}S(f)=\beta\log_{10}f+c,\qquad S(f)\propto f^\beta.
\end{equation}
A more negative slope indicates a more rapid decline in spectral power with frequency within the fitted band. A less negative slope implies relatively more high-frequency power, but does not necessarily imply a larger total variance. Agreement with the observed slope must be assessed over the same band. The $-5/3$ value is an inertial-subrange reference, not a universal target for a multiday wind-speed spectrum; a fitted slope near this value alone does not verify physically resolved turbulence.

\paragraph{Spectral error}
\begin{equation}
\varepsilon_S=\frac{\int |S_s(f)-S_o(f)|\,df}{\int S_o(f)\,df}.
\end{equation}
This dimensionless, normalized spectral distance is zero for identical spectra and increases as their levels or shapes differ. It is not bounded above by one. For example, a value of $0.5$ means that the integrated absolute spectral difference is half the integrated observed PSD over the comparison band; it does not mean a 50\% pointwise prediction error. Similar spectra can arise from differently timed signals, so spectral error should be considered together with correlation and coherence.

\paragraph{Coherence}
\begin{equation}
C_{os}(f)=\frac{|P_{os}(f)|^2}{P_{oo}(f)P_{ss}(f)},\qquad
\overline C=\langle C_{os}(f)\rangle_{f>0}.
\end{equation}
Magnitude-squared coherence ranges from zero to one and measures the strength of a linear frequency-specific relationship between the two signals. High coherence can coexist with a phase lag or amplitude mismatch; it does not imply identical signals. The reported average can conceal strong agreement at low frequencies and weak agreement at higher frequencies, and depends on spectral estimation and the included band. Thus, low band-averaged coherence can coexist with high time-domain correlation dominated by slowly varying weather events. Small differences in averaged coherence should not be treated as statistically significant without an uncertainty assessment.

\subsection{Spatial metrics}

Let $F_k(x,y)$ denote the horizontal wind-speed field at $10~\mathrm{m}$ in snapshot $k$. These diagnostics describe model structure rather than agreement with the two point anemometers; the domain-wide values are therefore repeated in both sensor tables.

\paragraph{Spatial variance}
\begin{equation}
V_{\mathrm{sp}}=\left\langle\left\langle
\bigl(F_k-\langle F_k\rangle_{x,y}\bigr)^2
\right\rangle_{x,y}\right\rangle_{\mathrm{frames}}.
\end{equation}
Larger values indicate stronger wind-speed contrasts across the domain at a given time, rather than stronger temporal fluctuations at a sensor. Zero corresponds to a spatially uniform speed field in every snapshot. Increased spatial variance may be relevant to differential plume advection, but cannot establish improved spatial accuracy without spatial observations.

\paragraph{Correlation length}
From the central-row profile of the normalized two-dimensional spatial autocorrelation, let $\ell_e$ denote the first index at which the correlation drops below $e^{-1}$ and $\ell_c$ the zero-lag index. Then
\begin{equation}
L=|\ell_e-\ell_c|\Delta x.
\end{equation}
Larger $L$ indicates that speed anomalies remain similar over greater distances along the sampled direction; smaller $L$ indicates more rapidly varying spatial structure. A central-row estimate is directional and does not characterize anisotropy in the full field. The identical $1000~\mathrm{m}$ values in the updated tables may reflect the available lag range; without confirming that the threshold is crossed, they should not be interpreted as evidence of identical physical correlation scales.

\paragraph{Spatial spectral peak}
\begin{equation}
k_{\mathrm{pk}}=\arg\max_{k>0}E(k),\qquad
\lambda_{\mathrm{pk}}=\frac{1}{k_{\mathrm{pk}}},\qquad
k=\sqrt{k_x^2+k_y^2}.
\end{equation}
Here $E(k)$ is the radially averaged two-dimensional power spectrum of the speed field and $k$ is measured in cycles per meter. If angular wavenumber is used instead, the wavelength conversion is $2\pi/k$. The peak identifies the strongest nonzero spatial spectral scale, not necessarily an individual eddy size. Peaks comparable to or larger than the domain width, as in the updated tables, can be controlled by the lowest wavenumber bins, domain-scale gradients, and boundary treatment; they should not be presented as well-resolved turbulence length scales.

\subsection{Dynamics metrics}

Using horizontal velocity components and finite differences on a grid with $\Delta x=\Delta y$, define vertical vorticity $\zeta$, horizontal divergence $\delta$, and the gradient diagnostic $\mathcal S$ by
\begin{equation}
\zeta=\frac{\partial v}{\partial x}-\frac{\partial u}{\partial y},\qquad
\delta=\frac{\partial u}{\partial x}+\frac{\partial v}{\partial y},\qquad
\mathcal S=\sqrt{\left(\frac{\partial u}{\partial x}\right)^2+
\left(\frac{\partial v}{\partial y}\right)^2}.
\end{equation}
All three have units of $\mathrm{s^{-1}}$. They are sensitive to grid resolution and differentiation, so comparisons should use consistent grids and numerical operators.

\paragraph{RMS vorticity}
\begin{equation}
\zeta_{\mathrm{rms}}=\frac{1}{N_f}\sum_{k=1}^{N_f}
\sqrt{\langle\zeta_k^2\rangle_{x,y}}.
\end{equation}
This metric measures typical rotational strength about the vertical axis without cancellation between opposite rotation signs. Larger values indicate stronger resolved horizontal velocity gradients associated with rotation, but do not establish that the vortices are physical or accurately located. The spatial RMS is taken before time averaging.

\paragraph{Maximum vorticity}
\begin{equation}
\zeta_{\max}=\max_k\max_{x,y}|\zeta_k|.
\end{equation}
The maximum emphasizes the strongest localized rotational event anywhere in the record. It is more sensitive than RMS vorticity to isolated extremes, boundary artifacts, grid spacing, and record length. It should not be interpreted as a measure of typical conditions or as inherently better when larger.

\paragraph{Enstrophy}
\begin{equation}
\mathcal E=\frac{1}{N_f}\sum_{k=1}^{N_f}
\frac{1}{2}\langle\zeta_k^2\rangle_{x,y}.
\end{equation}
This vertical-vorticity-based enstrophy has units of $\mathrm{s^{-2}}$ and measures mean squared rotational strength; it is not kinetic energy or the full three-dimensional enstrophy. Squaring makes it sensitive to strong gradients. Because RMS vorticity is averaged after taking a square root, the reported enstrophy need not equal one-half the square of the reported time-averaged RMS vorticity.

\paragraph{Mean divergence}
\begin{equation}
\overline\delta=\frac{1}{N_f}\sum_{k=1}^{N_f}
\langle\delta_k\rangle_{x,y}.
\end{equation}
Positive values indicate net horizontal spreading and negative values indicate net horizontal convergence. A value near zero can result from cancellation between strong local convergence and divergence, not an absence of either. Horizontal divergence at one height is not the full three-dimensional mass-continuity residual and should not be interpreted as a direct test of mass conservation.

\paragraph{RMS shear}
\begin{equation}
\mathcal S_{\mathrm{rms}}=\frac{1}{N_f}\sum_{k=1}^{N_f}
\sqrt{\langle\mathcal S_k^2\rangle_{x,y}}.
\end{equation}
For the definition given above, this quantity measures the magnitude of the horizontal normal velocity gradients. Despite the table label ``RMS shear,'' it does not include the cross derivatives $\partial u/\partial y$ and $\partial v/\partial x$, and is therefore not the full horizontal shear or strain-rate magnitude. The label is retained for consistency with the supplied tables; physical interpretation must follow the implemented derivatives. Larger values indicate stronger gradients under this definition, not automatically better plume-dispersion physics.

\subsection{Coherent-structure metrics}

\paragraph{POD mode energy}
For a temporally centered snapshot matrix $X\in\mathbb R^{n_t\times n_y n_x}$, let $\lambda_j$ denote its covariance eigenvalues, ordered from largest to smallest. The proper orthogonal decomposition (POD) fractions are
\begin{equation}
E_j=\frac{\lambda_j}{\sum_m\lambda_m},\qquad
E_{1:5}=\sum_{j=1}^{5}E_j.
\end{equation}
The first fraction measures how much spatial-field variance is represented by one coherent pattern; the cumulative fraction measures how much is captured by the first five patterns. Fractions near one indicate a highly concentrated, low-dimensional representation, whereas smaller values indicate variance distributed among more modes. For a speed-field decomposition, ``energy'' means variance of that input field, not the kinetic energy of all velocity components. If snapshots are not centered, the mean field can dominate the leading mode, so centering must be checked before interpreting the fractions as fluctuation variance. Neither a larger nor a smaller leading fraction is intrinsically more accurate.

\paragraph{Dominant DMD frequency}
Dynamic mode decomposition (DMD) describes snapshot evolution using modes with associated growth or decay rates and oscillation frequencies. For consecutive snapshots $X_1=[\mathbf x_0\cdots\mathbf x_{m-1}]$, $X_2=[\mathbf x_1\cdots\mathbf x_m]$, and rank-$r$ singular value decomposition $X_1=U_r\Sigma_rV_r^*$,
\begin{equation}
\widetilde A=U_r^*X_2V_r\Sigma_r^{-1},\qquad
\widetilde A\mathbf w_j=\mu_j\mathbf w_j,\qquad
\Phi=X_2V_r\Sigma_r^{-1}W.
\end{equation}
For snapshot interval $\Delta t_{\mathrm{snap}}$,
\begin{equation}
\omega_j=\frac{\ln\mu_j}{\Delta t_{\mathrm{snap}}},\qquad
f_j=\frac{|\Im\omega_j|}{2\pi},\qquad
\mathbf b=\Phi^+\mathbf x_0,
\end{equation}
\begin{equation}
f_{\mathrm{DMD}}=f_{j^\star},\qquad
j^\star=\arg\max_{j:f_j>0}|b_j|.
\end{equation}
The reported value selects the largest-amplitude nonzero-frequency mode under the chosen normalization, rather than a PSD peak or the mode with greatest integrated variance. Its period is $1/f_{\mathrm{DMD}}$; smaller frequencies indicate slower evolution. Results depend on rank truncation, centering, mode normalization, and snapshot spacing. A period approaching the full record length---as for $2.37\times10^{-6}~\mathrm{Hz}$, or approximately $4.9$ days, in the nudged two-way run---may represent slow drift or a poorly sampled oscillation rather than a reproducible periodic flow structure.

\bibliographystyle{plain}
\bibliography{references}

@Article{atmos13040510,
AUTHOR = {Chen, Qining and Modi, Mrinali and McGaughey, Gary and Kimura, Yosuke and McDonald-Buller, Elena and Allen, David T.},
TITLE = {Simulated Methane Emission Detection Capabilities of Continuous Monitoring Networks in an Oil and Gas Production Region},
JOURNAL = {Atmosphere},
VOLUME = {13},
YEAR = {2022},
NUMBER = {4},
ARTICLE-NUMBER = {510},
ISSN = {2073-4433}
}

@InProceedings{Fathi2023AGU,
  AUTHOR       = {Fathi, Arash and de Sousa Almeida, Joao Lucas and Bentivegna, Eloisa and Cardoso, Felipe and Elmegreen, Bruce and Klein, Levente and Mukkavilli, Karthik and Seastream, Grant and Sethuraman, Sandhya and Sundaram, Anantha and Trojak, Will},
  TITLE        = {Towards operational methane emission detection from oil and gas facilities through multi-modal sensing and advanced dispersion and atmospheric modeling},
  BOOKTITLE    = {AGU Fall Meeting 2023},
  YEAR         = {2023},
  ADDRESS      = {San Francisco, CA, USA},
  NOTE         = {Oral presentation},
  HOWPUBLISHED = {\url{https://research.ibm.com/publications/towards-operational-methane-emission-detection-from-oil-and-gas-facilities-through-multi-modal-sensing-and-advanced-dispersion-and-atmospheric-modeling}}
}

@InProceedings{Fathi2025AGU,
  AUTHOR       = {Fathi, Arash and de Sousa Almeida, Joao Lucas and Klein, Levente and Sundaram, Anantha and Zortea, Maciel},
  TITLE        = {Comparison of simulated and observed methane plumes at oil and gas sites in the {Permian Basin} using advanced dispersion, coupled mesoscale--{LES} atmospheric modeling, and scientific machine learning},
  BOOKTITLE    = {AGU Fall Meeting 2025},
  YEAR         = {2025},
  NOTE         = {Poster presentation},
  HOWPUBLISHED = {\url{https://research.ibm.com/publications/comparison-of-simulated-and-observed-methane-plumes-at-oil-and-gas-sites-in-the-permian-basin-using-advanced-dispersion-coupled-mesoscale-les-atmospheric-modeling-and-scientific-machine-learning}}
}

@InProceedings{Fathi2025AGUWind,
  AUTHOR       = {Fathi, Arash and Kharazmi, Ehsan and Menezes, Erin and Karniadakis, George Em and Sundaram, Anantha and Chen, Yuanlei},
  TITLE        = {High-resolution wind fields for operational methane monitoring at oil and gas sites by integrating physics-based simulations, field observations, and scientific machine learning},
  BOOKTITLE    = {AGU Fall Meeting 2025},
  YEAR         = {2025},
  VOLUME       = {442},
  NOTE         = {Abstract NG33B-0442}
}

@misc{wrf_wps_best_practices,
  author       = {{NCAR Mesoscale \& Microscale Meteorology Laboratory}},
  title        = {WRF Preprocessing System (WPS) Namelist Best Practices},
  year         = {2024},
  howpublished = {\url{https://www2.mmm.ucar.edu/wrf/users/namelist_best_prac_wps.html}},
  note         = {Accessed: 25 February 2026}
}

@article{Mirocha2010,
  author  = {Mirocha, J. D. and Lundquist, J. K. and Kosovi{\'c}, B.},
  title   = {Implementation of a Nonlinear Subfilter Turbulence Stress Model for Large-Eddy Simulation in the Advanced Research {WRF} Model},
  journal = {Monthly Weather Review},
  volume  = {138},
  pages   = {4212--4228},
  year    = {2010},
  doi     = {10.1175/2010MWR3286.1}
}

@article{Kosovic1997,
  author  = {Kosovi{\'c}, B.},
  title   = {Subgrid-Scale Modelling for the Large-Eddy Simulation of High-Reynolds-Number Boundary Layers},
  journal = {Journal of Fluid Mechanics},
  volume  = {336},
  pages   = {151--182},
  year    = {1997},
  doi     = {10.1017/S0022112096004697}
}

@article{Liu2020,
  author  = {Liu, Y. and Liu, Y. and Mu{\~n}oz-Esparza, D. and Hu, F. and Yan, C. and Miao, S.},
  title   = {Simulation of Flow Fields in Complex Terrain with {WRF--LES}: Sensitivity Assessment of Different {PBL} Treatments},
  journal = {Journal of Applied Meteorology and Climatology},
  volume  = {59},
  pages   = {1481--1501},
  year    = {2020},
  doi     = {10.1175/JAMC-D-19-0304.1}
}

@article{Talbot2012,
  author  = {Talbot, Charles and Bou-Zeid, Elie and Smith, Jim},
  title   = {Nested Mesoscale Large-Eddy Simulations with {WRF}: Performance in Real Test Cases},
  journal = {Journal of Hydrometeorology},
  volume  = {13},
  number  = {5},
  pages   = {1421--1441},
  year    = {2012},
  doi     = {10.1175/JHM-D-11-048.1}
}

@article{LiuNenes2012,
  author  = {Liu, P. and Tsimpidi, A. P. and Hu, Y. and Stone, B. and Russell, A. G. and Nenes, A.},
  title   = {Differences between Downscaling with Spectral and Grid Nudging Using {WRF}},
  journal = {Atmospheric Chemistry and Physics},
  volume  = {12},
  pages   = {3601--3610},
  year    = {2012},
  doi     = {10.5194/acp-12-3601-2012}
}

@article{Omrani2015,
  author  = {Omrani, Hiba and Drobinski, Philippe and Dubos, Thomas},
  title   = {Using Nudging to Improve Global--Regional Dynamic Consistency in Limited-Area Climate Modeling: What Should We Nudge?},
  journal = {Climate Dynamics},
  volume  = {44},
  pages   = {1627--1644},
  year    = {2015},
  doi     = {10.1007/s00382-014-2453-5}
}

@article{Daniels2016,
  author  = {Daniels, Megan H. and Lundquist, Katherine A. and Mirocha, Jeffrey D. and Wiersema, David J. and Chow, Fotini K.},
  title   = {A New Vertical Grid Nesting Capability in the Weather Research and Forecasting ({WRF}) Model},
  journal = {Monthly Weather Review},
  volume  = {144},
  number  = {10},
  pages   = {3725--3747},
  year    = {2016},
  doi     = {10.1175/MWR-D-16-0049.1}
}

@article{Bhimireddy2018,
  author  = {Bhimireddy, Sudheer R. and Bhaganagar, Kiran},
  title   = {Performance Assessment of Dynamic Downscaling of {WRF} to Simulate Convective Conditions during Sagebrush Phase 1 Tracer Experiments},
  journal = {Atmosphere},
  volume  = {9},
  number  = {12},
  pages   = {505},
  year    = {2018},
  doi     = {10.3390/atmos9120505}
}

@article{Ren2019,
  author  = {Ren, Hehe and Laima, Shujin and Chen, Wen-Li and Guo, Anxin and Li, Hui},
  title   = {Spatial Correlation-Based {WRF} Observation-Nudging Approach in Simulating Regional Wind Field},
  journal = {Wind and Structures},
  volume  = {28},
  number  = {2},
  pages   = {129--140},
  year    = {2019},
  doi     = {10.12989/was.2019.28.2.129}
}

@techreport{Haupt2019MMC,
  author      = {Haupt, S. E. and Allaerts, D. and Berg, L. and Churchfield, M. and DeCastro, A. and Draxl, C. and Gagne, D. J. and Hawbecker, P. and Jimenez, P. and Jonko, A. and Juliano, T. and Kaul, C. and Kosovi{\'c}, B. and McCandless, T. C. and Mirocha, J. and Mu{\~n}oz-Esparza, D. and Quon, E. and Rai, R. and Sauer, J. and Shaw, W.},
  title       = {{FY 2019 Report of the Atmosphere to Electrons Mesoscale-to-Microscale Coupling Project}},
  institution = {Pacific Northwest National Laboratory},
  number      = {PNNL-29603},
  address     = {Richland, Washington},
  month       = dec,
  year        = {2019},
  doi         = {10.2172/1735568}
}

@article{Haupt2023,
  author  = {Haupt, Sue Ellen and Kosovi{\'c}, Branko and Berg, Larry K. and Kaul, Colleen M. and Churchfield, Matthew and Mirocha, Jeffrey and Allaerts, Dries and Brummet, Thomas and Davis, Shannon and DeCastro, Amy and Dettling, Susan and Draxl, Caroline and Gagne, David John and Hawbecker, Patrick and Jha, Pankaj and Juliano, Timothy and Lassman, William and Quon, Eliot and Rai, Raj K. and Robinson, Michael and Shaw, William and Thedin, Regis},
  title   = {Lessons Learned in Coupling Atmospheric Models across Scales for Onshore and Offshore Wind Energy},
  journal = {Wind Energy Science},
  volume  = {8},
  pages   = {1251--1275},
  year    = {2023},
  doi     = {10.5194/wes-8-1251-2023}
}

@article{NayakKanda2023,
  author  = {Nayak, Sridhara and Kanda, Isao},
  title   = {Examining the Effectiveness of Doppler Lidar-Based Observation Nudging in {WRF} Simulation for Wind Field: A Case Study over Osaka, Japan},
  journal = {Atmosphere},
  volume  = {14},
  number  = {6},
  pages   = {972},
  year    = {2023},
  doi     = {10.3390/atmos14060972}
}

@article{GianiCrippa2024,
  author  = {Giani, Paolo and Crippa, Paola},
  title   = {On the Sensitivity of Large-Eddy Simulations of the Atmospheric Boundary Layer Coupled with Realistic Large-Scale Dynamics},
  journal = {Monthly Weather Review},
  volume  = {152},
  number  = {4},
  pages   = {1057--1075},
  year    = {2024},
  doi     = {10.1175/MWR-D-23-0101.1}
}

@article{AlOqaily2025,
  author  = {{Al Oqaily}, Dania and Giani, Paolo and Crippa, Paola},
  title   = {Evaluating {WRF} Multiscale Wind Simulations in Complex Terrain: Insights from the Perdig{\~a}o Field Campaign},
  journal = {Journal of Geophysical Research: Atmospheres},
  volume  = {130},
  pages   = {e2025JD044055},
  year    = {2025},
  doi     = {10.1029/2025JD044055}
}

@article{JaniszeskiCrippa2025,
  author  = {Janiszeski, Andrew and Crippa, Paola},
  title   = {Multiscale {WRF} Modeling of Meso- to Micro-Scale Flows during Sundowner Events},
  journal = {Journal of Geophysical Research: Atmospheres},
  volume  = {130},
  pages   = {e2024JD042972},
  year    = {2025},
  doi     = {10.1029/2024JD042972}
}

@article{Jiang2026,
  author  = {Jiang, Dong and Zhang, Qi and Hu, Qin and Wang, Zhifeng},
  title   = {Turbulent Design Parameters Simulation for Offshore Wind Turbines Combined {WRF--LES} Model and Observation Nudging Assimilation Method},
  journal = {Ocean Engineering},
  volume  = {354},
  pages   = {124933},
  year    = {2026},
  doi     = {10.1016/j.oceaneng.2026.124933}
}

@article{SkamarockKlemp2008,
  author  = {Skamarock, William C. and Klemp, Joseph B.},
  title   = {A Time-Split Nonhydrostatic Atmospheric Model for Weather Research and Forecasting Applications},
  journal = {Journal of Computational Physics},
  volume  = {227},
  number  = {7},
  pages   = {3465--3485},
  year    = {2008},
  doi     = {10.1016/j.jcp.2007.01.037}
}

@article{Wyngaard2004,
  author  = {Wyngaard, John C.},
  title   = {Toward Numerical Modeling in the {``Terra Incognita''}},
  journal = {Journal of the Atmospheric Sciences},
  volume  = {61},
  number  = {14},
  pages   = {1816--1826},
  year    = {2004},
  doi     = {10.1175/1520-0469(2004)061<1816:TNMITT>2.0.CO;2}
}

@article{Hong2006,
  author  = {Hong, Song-You and Noh, Yign and Dudhia, Jimy},
  title   = {A New Vertical Diffusion Package with an Explicit Treatment of Entrainment Processes},
  journal = {Monthly Weather Review},
  volume  = {134},
  number  = {9},
  pages   = {2318--2341},
  year    = {2006},
  doi     = {10.1175/MWR3199.1}
}

@article{Moeng2007,
  author  = {Moeng, C.-H. and Dudhia, Jimy and Klemp, Joe and Sullivan, Peter},
  title   = {Examining Two-Way Grid Nesting for Large Eddy Simulation of the {PBL} Using the {WRF} Model},
  journal = {Monthly Weather Review},
  volume  = {135},
  number  = {6},
  pages   = {2295--2311},
  year    = {2007},
  doi     = {10.1175/MWR3406.1}
}

@article{Mirocha2013,
  author  = {Mirocha, Jeff and Kirkil, Gokhan and Bou-Zeid, Elie and Chow, Fotini Katopodes and Kosovi{\'c}, Branko},
  title   = {Transition and Equilibration of Neutral Atmospheric Boundary Layer Flow in One-Way Nested Large-Eddy Simulations Using the {Weather Research and Forecasting} Model},
  journal = {Monthly Weather Review},
  volume  = {141},
  number  = {3},
  pages   = {918--940},
  year    = {2013},
  doi     = {10.1175/MWR-D-11-00263.1}
}

@article{MunozEsparza2014,
  author  = {Mu{\~n}oz-Esparza, Domingo and Kosovi{\'c}, Branko and Mirocha, Jeff and van Beeck, Jeroen},
  title   = {Bridging the Transition from Mesoscale to Microscale Turbulence in Numerical Weather Prediction Models},
  journal = {Boundary-Layer Meteorology},
  volume  = {153},
  number  = {3},
  pages   = {409--440},
  year    = {2014},
  doi     = {10.1007/s10546-014-9956-9}
}

@article{Skamarock2004,
  author  = {Skamarock, William C.},
  title   = {Evaluating Mesoscale {NWP} Models Using Kinetic Energy Spectra},
  journal = {Monthly Weather Review},
  volume  = {132},
  number  = {12},
  pages   = {3019--3032},
  year    = {2004},
  doi     = {10.1175/MWR2830.1}
}

@article{StaufferSeaman1994,
  author  = {Stauffer, David R. and Seaman, Nelson L.},
  title   = {Multiscale Four-Dimensional Data Assimilation},
  journal = {Journal of Applied Meteorology},
  volume  = {33},
  number  = {3},
  pages   = {416--434},
  year    = {1994},
  doi     = {10.1175/1520-0450(1994)033<0416:MFDDA>2.0.CO;2}
}

@article{Sommerfeld2019,
  author  = {Sommerfeld, Markus and D{\"o}renk{\"a}mper, Martin and Steinfeld, Gerald and Crawford, Curran},
  title   = {Improving Mesoscale Wind Speed Forecasts Using Lidar-Based Observation Nudging for Airborne Wind Energy Systems},
  journal = {Wind Energy Science},
  volume  = {4},
  number  = {4},
  pages   = {563--580},
  year    = {2019},
  doi     = {10.5194/wes-4-563-2019}
}

@article{Yi2020,
  author  = {Yi, Xue and Li, Deqin and Zhao, Chunyu and Shen, Lidu and Zhou, Xiaoyu},
  title   = {Impact of a Dense Surface Network on High-Resolution Dynamical Downscaling via Observation Nudging},
  journal = {Journal of Applied Meteorology and Climatology},
  volume  = {59},
  number  = {10},
  pages   = {1655--1670},
  year    = {2020},
  doi     = {10.1175/JAMC-D-20-0071.1}
}

@article{Li2016Nudging,
  author  = {Li, Xiangshang and Choi, Yunsoo and Czader, Beata and Roy, Anirban and Kim, Hyuncheol and Lefer, Barry and Pan, Shuai},
  title   = {The Impact of Observation Nudging on Simulated Meteorology and Ozone Concentrations during {DISCOVER-AQ} 2013 {Texas} Campaign},
  journal = {Atmospheric Chemistry and Physics},
  volume  = {16},
  number  = {5},
  pages   = {3127--3144},
  year    = {2016},
  doi     = {10.5194/acp-16-3127-2016}
}

\end{document}